\documentclass[aps, prd, superscriptaddress, preprintnumbers]{revtex4}

\usepackage{graphicx,amsfonts,amsmath,amssymb,amstext}
\usepackage{float,wrapfig}
\usepackage{subfigure, psfrag}
\usepackage{dsfont}
\usepackage{color}
\usepackage{bm}

\newcommand{\be}{\begin{equation}}
\newcommand{\ee}{\end{equation}}
\newcommand{\ba}{\begin{eqnarray}}
\newcommand{\ea}{\end{eqnarray}}

\definecolor{purple}{rgb}{0.8,0,0.6}
\definecolor{darkgreen}{rgb}{0.00,0.6,0.00}
\begin{document}

\title{Hydrodynamic electron flow in a Weyl semimetal slab: Role of Chern--Simons terms}
\date{May 14, 2017}

\author{E.~V.~Gorbar}
%\email{gorbar@bitp.kiev.ua}
\affiliation{Department of Physics, Taras Shevchenko National Kiev University, Kiev, 03680, Ukraine}
\affiliation{Bogolyubov Institute for Theoretical Physics, Kiev, 03680, Ukraine}

\author{V.~A.~Miransky}
%\email{vmiransk@uwo.ca}
\affiliation{Department of Applied Mathematics, Western University, London, Ontario, Canada N6A 5B7}

\author{I.~A.~Shovkovy}
%\email{igor.shovkovy@asu.edu}
\affiliation{College of Integrative Sciences and Arts, Arizona State University, Mesa, Arizona 85212, USA}
\affiliation{Department of Physics, Arizona State University, Tempe, Arizona 85287, USA}

\author{P.~O.~Sukhachov}
%\email{psukhach@uwo.ca}
\affiliation{Department of Applied Mathematics, Western University, London, Ontario, Canada N6A 5B7}

\begin{abstract}
The hydrodynamic flow of the chiral electron fluid in a Weyl semimetal slab of finite thickness is studied
by using the consistent hydrodynamic theory. The latter includes viscous, anomalous, and vortical effects, as
well as accounts for dynamical electromagnetism. The energy and momentum separations between the
Weyl nodes are taken into account via the topological Chern--Simons contributions in the electric current
and charge densities in Maxwell's equations. When an external electric field is applied parallel
to the slab, it is found that the electron fluid velocity has a nonuniform profile determined by the
viscosity and the no-slip boundary conditions. Most remarkably, the fluid velocity field develops a nonzero
component across the slab that gradually dissipates when approaching the surfaces. This abnormal
component of the flow arises due to the anomalous Hall voltage induced by the topological
Chern--Simons current. Another signature feature of the hydrodynamics in Weyl semimetals is a strong
modification of the anomalous Hall current along the slab in the direction perpendicular to the
applied electric field. Additionally, it is found that the topological current induces an electric
potential difference between the surfaces of the slab that is strongly affected by the hydrodynamic flow.
\end{abstract}

\maketitle

\section{Introduction}
\label{sec:Introduction}

Charge transport in solids in one of the most basic and paradigmatic phenomena in condensed matter
physics. Usually, it could be described qualitatively by the Drude model, which assumes that individual electrons
are accelerated by an applied electric field and slowed down by collisions with impurities and/or
phonons. In the 1960s, however, Gurzhi proposed \cite{Gurzhi,Gurzhi-effect} that a qualitatively different hydrodynamic regime
of the electron transport could be possible if the electron-electron scattering rate dominates over the electron-impurity
and electron-phonon ones. In such a case, the electrons do not move independently but participate in a collective
motion as a liquid. Since the electron-impurity scattering rate increases as temperature decreases and the
electron-phonon interaction rate becomes large at high temperatures, the electron hydrodynamic transport
cannot be easily realized in most metals.

For the first time, a hydrodynamic electron flow in a condensed matter system was experimentally observed
in a two-dimensional (2D) electron gas of high-mobility $\mbox{(Al,Ga)As}$ heterostructures in the 1990s
\cite{Molenkamp,Jong-Molenkamp:1995}. Recently, a large viscous contribution to the resistance of the
ultrapure 2D metal palladium cobaltate ($\mbox{PdCoO}_2$) was reported in Ref.~\cite{Moll}. The
hydrodynamic transport was also confirmed in graphene~\cite{Crossno,Ghahari}.
The conditions of its realization
and the viscosity effects were theoretically studied in
Refs.~\cite{Torre-Polini:2015,Pellegrino-Polini:2016,Levitov-Falkovich:2016,Alekseev-Dmitriev:2017,Levitov:2017,Ho-Adam:2017,Falkovich-Levitov-PRL:2017} (for a recent review, see Ref.~\cite{Lucas:2017idv}).
Among these effects, one could identify nonlocal transport affected by the current vortices
\cite{Torre-Polini:2015,Pellegrino-Polini:2016,Levitov-Falkovich:2016,Falkovich-Levitov-PRL:2017}
and a higher than ballistic conduction in a graphene constriction \cite{Levitov:2017}. The Dyakonov--Shur
instability \cite{Dyakonov-Shur:1993} and a frequency-dependent response in the Corbino disk
\cite{Tomadin-Polini-PRL:2014} also allow one to quantify the electron viscosity.

In 2017, the hydrodynamic
regime was experimentally observed in tungsten diphosphide ($\mbox{WP}_2$) \cite{Gooth:2017},
which is a Weyl semimetal and, unlike earlier materials, is a three-dimensional (3D) material. The observation
of the hydrodynamic regime in $\mbox{WP}_2$ is supported by a characteristic dependence of the electrical resistivity on
the cross section of the constriction channel, as well as a strong violation of the Wiedemann--Franz
law with the lowest value of the Lorenz number ever reported.

Weyl semimetals are unusual materials whose low-energy excitations are described by the
relativistic-like Weyl equation. As proved by Nielsen and Ninomiya \cite{Nielsen-Ninomiya,Nielsen-Ninomiya-2}, the Weyl nodes
in lattice systems always come in pairs of opposite chirality. Many real materials usually have multiple pairs
of opposite chirality nodes (for recent reviews, see Refs.~\cite{Yan-Felser:2017-Rev,Hasan-Huang:2017-Rev,Armitage-Vishwanath:2017-Rev}).
In each pair, the nodes are separated by $2\mathbf{b}$ in momentum and/or  $2b_0$ in
energy. (The parameter $\mathbf{b}$ is known as the chiral shift \cite{Gorbar:2009bm}.)
When the sums of all chiral shifts and/or energy separations are nonzero,
the time-reversal (TR) and/or parity-inversion (PI) symmetries, respectively, are broken.

From a theoretical viewpoint, it is important that the Weyl nodes are monopoles of the Berry curvature
\cite{Berry:1984} with nonzero topological charges. The chiral nature of quasiparticles as well as the underlying
topology affect the physical properties of Weyl semimetals. One of the signature observables is the negative longitudinal
(with respect to the direction of an external magnetic field) magnetoresistivity \cite{Nielsen,Son:2012bg}.
(For recent reviews of the transport phenomena in Weyl semimetals, see Refs.~\cite{Lu-Shen-rev:2017,Wang-Lin-rev:2017,Gorbar:2017lnp}.)

The transport properties of Weyl materials are usually theoretically studied by using the Kubo formalism or
semiclassical methods such as the chiral kinetic theory \cite{Son:2012wh,Stephanov:2012ki,Hidaka:2016yjf,Hidaka:2017auj}
and the chiral hydrodynamics \cite{Son:2009tf,Sadofyev:2010pr,Neiman:2010zi}.
In particular, the hydrodynamic equations are a standard means for
studying interacting systems close to equilibrium in the long-wavelength limit and at large timescales \cite{Landau:t6}.
The formulation of the hydrodynamic framework for chiral
plasmas was proposed in Refs.~\cite{Son:2009tf,Sadofyev:2010pr,Neiman:2010zi}. In addition to the space-time
evolution of conserved quantities, such as energy, momentum, and electric charge, it includes
the chiral charge, whose conservation is violated {only} by the chiral anomaly \cite{Adler,Bell-Jackiw}.
As expected, the chiral hydrodynamic
approach can describe the negative magnetoresistance \cite{Landsteiner:2014vua,Lucas:2016omy} and the
thermoelectric transport \cite{Lucas:2016omy} in Weyl semimetals. However, the corresponding equations lack
any information on the separation between the Weyl nodes.
Recently, we argued \cite{Gorbar:2017vph} that, as in the consistent chiral kinetic theory \cite{Gorbar:2016ygi}, the topological Chern--Simons contributions (also known as the Bardeen--Zumino terms in high-energy physics \cite{Bardeen}) should be added to the electric current and charge densities in Maxwell's equations of the \emph{consistent hydrodynamics} (CHD).
(As demonstrated by the analysis in a two-band model of a Weyl semimetal \cite{Gorbar:2017wpi},
the microscopic origin of the topological Chern--Simons contributions is related
to the filled electron states deep below the Fermi surface.)
These contributions introduce the missing dependence on the energy separation $b_0$
and the chiral shift $\mathbf{b}$ that plays a critical role in reproducing the correct chiral magnetic effect
(CME) \cite{Franz:2013,Basar:2014,Landsteiner:2016} and the anomalous Hall effect (AHE)
\cite{Ran,Burkov:2011ene,Grushin-AHE,Goswami,Burkov-AHE:2014} in Weyl semimetals. Note that the former vanishes in the equilibrium
state of a Weyl semimetal as required by general principles \cite{Franz:2013}.

As we showed in Ref.~\cite{Gorbar:2017vph}, the Chern--Simons terms do not enter directly the Euler
equation and the energy conservation relation. However, the hydrodynamics of the charged electron fluid is still affected
by such topological terms via a self-consistent treatment of dynamical electromagnetic fields. Indeed, by
making use of the CHD, we found \cite{Gorbar:2017vph,Gorbar:2018nmg} that the Chern--Simons
contributions strongly modify the dispersion relations of collective modes in Weyl semimetals. One of the key
predictions, in particular, is the existence of distinctive anomalous Hall waves sustained by the
local AHE currents.

In this paper, by making use of the CHD, we study a hydrodynamic flow of chiral electrons
in a Weyl semimetal with a simple slab geometry, i.e., a finite-thickness slab that is infinite in two
directions.
Our principal finding is that the Chern--Simons currents indeed influence the electron fluid leading to a distinctive normal flow across the slab.
The latter affects the electric current in the direction parallel to the slab surfaces but perpendicular to the applied electric field.
Additionally, the flow modifies an electric potential difference between the surfaces of the slab.

The paper is organized as follows. In Sec.~\ref{sec:model} we review the basic features
of the CHD in Weyl semimetals \cite{Gorbar:2017vph} and introduce additional terms related to the
viscosity and the intrinsic conductivity. The model setup, the boundary conditions, and the linearized CHD
equations are given in Sec.~\ref{sec:LCHD-general}. Sections~\ref{sec:LCHD-hydro-flows-PI} and
\ref{sec:LCHD-hydro-flows-PI-broken} are devoted to the hydrodynamic electron transport in Weyl
semimetals with intact and broken PI symmetries, respectively. Our results are summarized and discussed in Sec.~\ref{sec:Summary}.
Some technical details, including the general form of the CHD equations as well as
additional hydrodynamic variables, are given in Appendices~\ref{sec:app-CHD-eqs}
and \ref{sec:app-hydro-vars}. Throughout this paper, we set the Boltzmann constant $k_B=1$.

\section{Steady states in consistent hydrodynamic theory}
\label{sec:model}

In this section, we briefly discuss the key features of the steady-state CHD equations in Weyl and Dirac semimetals
derived in Ref.~\cite{Gorbar:2017vph} and amended here by the viscosity of the electron fluid as well as the intrinsic Ohmic
contributions to the currents.
The explicit
form of the continuity relations for the electric and chiral charges, the Navier--Stokes equation, and the energy
conservation relation is given in Appendix~\ref{sec:app-CHD-eqs}.

It is instructive to emphasize that the electric $\mathbf{J}$ and chiral $\mathbf{J}_5$ current densities in Weyl semimetals, which
satisfy the standard continuity relations
(\ref{model-J-conserv-eq}) and (\ref{model-J5-conserv-eq}) in Appendix \ref{sec:app-CHD-eqs},
are the \emph{total} ones including both material and topological terms.
Their explicit expressions in the presence of electric $\mathbf{E}$ and magnetic $\mathbf{B}$ fields read
\begin{eqnarray}
\label{model-J-def}
\mathbf{J} &=& -en\mathbf{u}+\sigma\mathbf{E}+\kappa_e\bm{\nabla}T +\frac{\sigma_5}{e} \bm{\nabla}\mu_5
+\sigma^{(V)}\bm{\omega} +\sigma^{(B)}\mathbf{B}+\frac{c\sigma^{(B)}\left[\mathbf{E}\times\mathbf{u}\right]}{3 v_F^2}  + \frac{5c^2\sigma^{(\epsilon, u)}\left[\mathbf{E}\times\bm{\omega}\right]}{v_F^2} -\frac{\left[\mathbf{u}\times\bm{\nabla}\right]\sigma^{(V)}}{3}
 \nonumber\\
&+&\frac{\left[\bm{\nabla}\times\bm{\omega}\right] \sigma^{(\epsilon, V)}}{2}+\mathbf{J}_{\text{{\tiny CS}}},\\
\label{model-J5-def}
\mathbf{J}_5 &=& -en_{5}\mathbf{u}+\sigma_5\mathbf{E}+\kappa_{e,5}\bm{\nabla}T +\frac{\sigma}{e} \bm{\nabla}\mu_5
+\sigma^{(V)}_5\bm{\omega} +\sigma^{(B)}_5\mathbf{B}+\frac{c\sigma^{(B)}_5\left[\mathbf{E}\times\mathbf{u}\right]}{3v_F^2} -\frac{\left[\mathbf{u}\times\bm{\nabla}\right]\sigma^{(V)}_5}{3} +\frac{\left[\bm{\nabla}\times\bm{\omega}\right] \sigma^{(\epsilon, V)}_5}{2}.
\end{eqnarray}
Here $e$ is the absolute value of the electron charge, $\mathbf{u}$ is the electron fluid velocity, $-en$ and $-en_{5}$ are the matter parts of
the electric and chiral charge densities (defined in the absence of electromagnetic fields and the fluid velocity),
respectively, $\mu_5$ is the chiral chemical potential, $T$ is temperature, $\bm{\omega}=\left[\bm{\nabla}\times\mathbf{u}\right]/2$
is the vorticity, $v_F$ is the Fermi velocity, and $c$ is the speed of light.

Compared to the currents used in Ref.~\cite{Gorbar:2017vph}, Eqs.~(\ref{model-J-def})
and (\ref{model-J5-def}) contain a few new terms
related to the intrinsic (also known in the holography framework as quantum critical or incoherent
\cite{Hartnoll:2007ih,Kovtun:2008kx,Hartnoll:2014lpa,Davison:2015taa,Lucas:2015lna}) electric $\sigma$ and thermoelectric $\kappa_e$ conductivities, as well as their chiral
counterparts $\sigma_5$ and $\kappa_{e,5}$, respectively.
The intrinsic electric conductivity $\sigma$ is extensively discussed
in the holographic approach (see, e.g., Refs.~\cite{Hartnoll:2007ih,Kovtun:2008kx,Landsteiner:2014vua,Hartnoll:2014lpa,Davison:2015taa,Lucas:2015lna,Lucas:2017idv})
and is related to the nonhydrodynamic part of the distribution function.
It was shown that $\sigma$ is nonzero even in clean samples at the
neutrality point, i.e., at vanishing electric $\mu$ and chiral $\mu_5$ chemical potentials.
In addition, the intrinsic electric conductivity \cite{Holder-Ostrovsky:2017} could originate from the
nonperturbative rare-region effects \cite{Nandkishore:2014} and charge puddles \cite{Rodionov:2015} in dirty Weyl semimetals.
In general, $\sigma$ depends on the thermodynamic variables, i.e., $\mu$, $\mu_5$, and $T$.
Unfortunately, its explicit form is not universal and cannot be easily fixed.
In our study, we use the intrinsic conductivity similar to that obtained in the holographic approach in Refs.~\cite{Hartnoll:2007ih,Kovtun:2008kx,Hartnoll:2014lpa,Landsteiner:2014vua,Davison:2015taa,Lucas:2015lna}, i.e.,
\begin{equation}
\label{model-sigma}
\sigma = \frac{3 \pi^2\hbar v_F^3}{2\pi} \left(\frac{\partial n}{\partial \mu} +\frac{\partial n_5}{\partial \mu_5}\right)
\tau_{ee},
\end{equation}
where $\tau_{ee}$ is the electron-electron scattering rate. The latter is a very important characteristics
because the hydrodynamic regime could be realized in a Weyl semimetal only when $\tau_{ee}$
is sufficiently small. According to Ref.~\cite{Gooth:2017}, the experimental value of $\tau_{ee}$ in
the hydrodynamic regime of WP$_2$ is well approximated by the expression $\hbar/T$ even though the electric chemical potential
is much larger than temperature. Such an estimate for $\tau_{ee}$ suggests that the electron fluid in WP$_2$ is
strongly interacting and cannot be described by a conventional Fermi liquid. Indeed, if the electrons in WP$_2$
formed a Fermi liquid, the quasiparticle lifetime due to the electron-electron interactions would be $\tau_{ee} \sim\hbar \mu /\left(\alpha T\right)^2$, where $\alpha$ is the coupling constant. Such a lifetime is parametrically much larger than the experimentally extracted one.
If $\tau_{ee}$ were consistent with the Fermi liquid expression,
the hydrodynamic effects observed in Ref.~\cite{Gooth:2017} would not be possible.
In our study, therefore, we use $\tau_{ee}=\hbar/T$ which is consistent with the findings in Ref.~\cite{Gooth:2017}.

To the best of our knowledge, the intrinsic chiral conductivity $\sigma_5$ is not studied in the literature.
We believe that the corresponding expression could be obtained in a similar way to Eq.~(\ref{model-sigma}) and is given by
\begin{equation}
\label{model-sigma5}
\sigma_5 = \frac{3 \pi^2\hbar v_F^3}{2\pi} \left(\frac{\partial n_5}{\partial \mu}+\frac{\partial n}{\partial \mu_5}\right) \tau_{ee}.
\end{equation}

By following the standard approach (see, e.g., Ref.~\cite{Lucas:2017idv}) we also included the intrinsic thermoelectric currents $\kappa_e\bm{\nabla}T$ and
$\kappa_{e,5}\bm{\nabla}T$ in the electric (\ref{model-J-def}) and chiral (\ref{model-J5-def}) current densities.
The corresponding coefficients are proportional to the intrinsic conductivities and
the chemical potentials, i.e.,
\begin{eqnarray}
\label{model-kappae}
\kappa_e=-\frac{\mu \sigma +\mu_5 \sigma_5}{eT},\\
\label{model-kappae5}
\kappa_{e,5}=-\frac{\mu \sigma_5 +\mu_5 \sigma}{eT}.
\end{eqnarray}
Similarly to the thermoelectric currents, we also added the terms related to the gradient of the chiral chemical potential $\propto \bm{\nabla}\mu_5$ in Eqs.~(\ref{model-J-def}) and (\ref{model-J5-def}).

One of the defining features of the CHD is the presence of the topological Chern--Simons (or Bardeen--Zumino \cite{Bardeen})
contributions~\cite{Landsteiner:2013sja,Landsteiner:2016,Gorbar:2016ygi,Gorbar:2017wpi,Gorbar:2017vph}.
In components, the Chern--Simons terms take the following form:
\begin{eqnarray}
\rho_{\text{{\tiny CS}}} &=&- \frac{e^3 (\mathbf{b}\cdot\mathbf{B})}{2\pi^2\hbar^2c^2},
\label{model-CS-charge}
 \\
\mathbf{J}_{\text{{\tiny CS}}} &=&-\frac{e^3b_0 \mathbf{B}}{2\pi^2\hbar^2c} + \frac{e^3\left[\mathbf{b}\times\mathbf{E}\right]}{2\pi^2\hbar^2c},
\label{model-CS-current}
\end{eqnarray}
where $b_0$ and $\mathbf{b}$ correspond to the energy and momentum separations
between the Weyl nodes. The Chern--Simons term in the charge density (\ref{model-CS-charge})
is not directly involved in the equations describing the steady-state
hydrodynamics. However, as we will show below, this term plays an important
role in determining the equilibrium electric chemical potential
when $\mathbf{b}\cdot\mathbf{B}\neq0$.
Further, the anomalous transport coefficients, including $\sigma^{(B)}$, $\sigma^{(\epsilon, u)}$,
etc., are defined in Eqs.~(\ref{model-sigma-CKT-be})--(\ref{model-sigma-CKT-ee}) in Appendix \ref{sec:app-CHD-eqs}. It is worth noting that in Weyl semimetals with multiple pairs of Weyl nodes the chiral shift should be replaced with an effective one, $\mathbf{b}\to \mathbf{b}_{\rm eff}=\sum_{n}\mathbf{b}^{(n)}$, where the sum runs over all pairs of Weyl nodes. Therefore, in order to obtain a nonzero AHE current, captured by the last term in Eq.~(\ref{model-CS-current}), the sum should be nonvanishing $\sum_{n}\mathbf{b}^{(n)}\neq\mathbf{0}$. The latter is indeed the case when the TR symmetry is broken.

As expected, the expressions for the current densities in Eqs.~(\ref{model-J-def}) and (\ref{model-J5-def})
include the usual terms associated with the conventional CME \cite{Kharzeev:2007tn,Kharzeev:2007jp,Fukushima:2008xe}
and the chiral vortical effect (CVE) \cite{Chen:2014cla}, as well as the chiral separation effects driven by the
magnetic field and vorticity. They also contain contributions due to the simultaneous presence of the electric field
and the fluid velocity/vorticity, as well as the terms proportional to the cross product of the fluid velocity/vorticity
and the gradients of the thermodynamic variables.
In addition, unlike the chiral current density $\mathbf{J}_5$, the electric one $\mathbf{J}$ contains the term $\propto [\mathbf{E}\times\bm{\omega}]$, which resembles the
anomalous inflow term $\propto [\mathbf{E}\times\mathbf{u}]$, albeit does not depend on the chemical potentials of the system.

It is well known that the hydrodynamic equations can be obtained from the kinetic ones by averaging the latter
with the quasiparticle momentum and energy \cite{Landau:t10,Huang-book}. The corresponding Euler equation
and the energy conservation relation in the CHD without viscosity were obtained in Ref.~\cite{Gorbar:2017vph}.
The Euler equation in viscous fluids should be amended by the following standard terms (see, e.g.,
Ref.~\cite{Landau:t6}):
\begin{equation}
\label{model-D}
\eta \Delta \mathbf{u} +\left(\zeta+\frac{\eta}{3}\right) \bm{\nabla}\left(\bm{\nabla}\cdot\mathbf{u}\right),
\end{equation}
where $\eta$ and $\zeta$ denote the shear and bulk dynamic viscosities, respectively.
In relativistic-like systems, $\eta = \eta_{\rm kin}(\epsilon+P)/v_F^2$, where $\eta_{\rm kin}\approx v_F^2\tau_{ee}/4$ is the kinematic shear viscosity (see, e.g., Ref.~\cite{Alekseev:2016}) and $\zeta=0$ \cite{Landau:t10}.
In our study, for the sake of simplicity, we ignore the spatial dependence of the viscosity coefficients.
The resulting Navier--Stokes equation is given by Eq.~(\ref{model-Euler}) in Appendix \ref{sec:app-CHD-eqs}. Further,
the energy dissipation terms due to viscosity and thermoconductivity should be also included in the energy conservation
relation \cite{Landau:t6}, i.e.,
\begin{equation}
\label{model-D0}
\eta \left(\mathbf{u}\Delta \mathbf{u}\right) +\left(\zeta+\frac{\eta}{3}\right) \left(\mathbf{u}\cdot\bm{\nabla}\right)\left(\bm{\nabla}\cdot\mathbf{u}\right) + \kappa \Delta T,
\end{equation}
where we also assumed that $\kappa$ is uniform.
The complete form of the corresponding relation is given by Eq.~(\ref{model-energy}) in Appendix \ref{sec:app-CHD-eqs}.
The last term in Eq.~(\ref{model-D0}) is related to the thermoconductivity and, as we will see below,
is important for the self-consistency of the CHD equations
describing the flow of the electron fluid. The corresponding coefficient can be approximated by
\begin{equation}
\label{model-kappa}
\kappa = \kappa_0 \frac{\pi^2 T}{3e^2} \sigma_{\rm exp},
\end{equation}
where $\kappa_0\lesssim 1$ quantifies the deviation from the Wiedemann--Franz law and $\sigma_{\rm exp}$ is the total electric conductivity measured experimentally.
It is important to emphasize that both the Navier--Stokes equation (\ref{model-Euler}) and the energy continuity relation
(\ref{model-energy})
do not depend on the Chern--Simons current density $\mathbf{J}_{\text{\tiny CS}}$ \cite{Gorbar:2017wpi,Gorbar:2017vph}.
In addition, they do not contain the intrinsic Ohmic $\sigma \mathbf{E}$ and thermoelectric $\kappa_e \bm{\nabla}T$ current densities, as well as the gradient term $(\sigma_5/e) \bm{\nabla}\mu_5$.
However, the hydrodynamic equations of the charged fluid (\ref{model-J-conserv-eq}), (\ref{model-J5-conserv-eq}),
(\ref{model-Euler}), and (\ref{model-energy}) should be supplemented by the steady-state
Maxwell's equations
\begin{eqnarray}
\label{model-Maxwell-be}
&\varepsilon_e\bm{\nabla}\cdot\mathbf{E} = 4\pi (\rho+\rho_{\rm b}), \qquad & \bm{\nabla}\times\mathbf{E} =0, \\
&\bm{\nabla}\times\mathbf{B} = \mu_m \frac{4\pi}{c}\mathbf{J}, \qquad  & \bm{\nabla}\cdot\mathbf{B} = 0,
\label{model-Maxwell-ee}
\end{eqnarray}
where $\varepsilon_e$ and $\mu_m$ denote background electric permittivity and magnetic permeability, respectively.
Maxwell's equations contain the total electric current density $\mathbf{J}$ given in
Eq.~(\ref{model-J-def}) and the {\em total} electric charge density of electrons $\rho$ obtained in
Ref.~\cite{Gorbar:2017vph}. The latter reads
\begin{equation}
\label{model-rho-def}
\rho = -e n +\frac{\sigma^{(B)} \left(\mathbf{B}\cdot\mathbf{u}\right)}{3v_F^2}
+ \frac{5c \sigma^{(\epsilon, u)} \left(\mathbf{B}\cdot\bm{\omega}\right)}{v_F^2} +\rho_{\text{{\tiny CS}}}.
\end{equation}
We emphasize that the electric charge density in Gauss's law also includes the background
contribution $\rho_b$ due to the electrons in the inner shells and the ions of the lattice.

Before finishing this section, let us note that the CHD in this study treats the
left- and right-handed quasiparticles as parts of a single fluid with a common flow velocity $\mathbf{u}$.
(For the discussion of single- and two-fluid approaches in various types of plasmas, see
Refs.~\cite{Gross-Krook:1956,Braginskii:1965}.) In this connection, a few remarks are in order.
First, we note that unlike the electron-hole plasma in graphene, where the two-fluid approach
was used at small charge densities \cite{Foster-Aleiner:2009,Alekseev-Dmitriev:2017,Svintsov:2018},
the electric chemical potential is usually rather large in realistic Weyl semimetals. Further, according to
Refs.~\cite{Gross-Krook:1956,Braginskii:1965}, a two-fluid description might be needed only
when the inverse relaxation time of the interfluid perturbations is comparable to or smaller than the
frequencies of processes inducing the separation of the fluid components.  Obviously, this is
not the case in the steady-state regime under consideration. Therefore, we treat the hydrodynamic
electric charge transport by using the single-fluid description. In general, the use of two separate chiral
fluids could be also essential when there are chirality-dependent forces. However, as we will show
below, such forces are absent in the linearized version of the Navier--Stokes equations; see
Eqs.~(\ref{LCHD-mu5-Euler-x})--(\ref{LCHD-mu5-Euler-z}). (The possibility of the chiral
fluids splitting will be further discussed at the end of Sec.~\ref{sec:Summary}.)

\section{Linearized CHD and model setup}
\label{sec:LCHD-general}

In this section, we define the model setup and present the linearized CHD equations with the appropriate
boundary conditions. Before discussing the specific details, we note that in the hydrodynamic
regime all thermodynamic variables and electromagnetic fields may deviate from their global equilibrium values and, in general, depend on the spatial coordinates and time.
In this study, however, we will limit ourselves only to steady states in which there is no time dependence, i.e.,
$\mu=\mu(\mathbf{r})$, $\mu_5=\mu_5(\mathbf{r})$, etc.

\subsection{Equilibrium state}
\label{sec:LCHD-general-eq-vars}

It is instructive to start our discussion with the definition of the thermodynamic variables in the global equilibrium state when the background
electromagnetic fields vanish. In this case, the energy density, the pressure, as well as the electric and chiral
charge densities can be given in terms of temperature and the chemical potentials:
\begin{eqnarray}
\label{LCHD-general-equilibrium-be}
\epsilon &=& \frac{\mu^4+6\mu^2\mu_{5}^2+\mu_{5}^4}{4\pi^2\hbar^3v_F^3}
+\frac{T^2(\mu^2+\mu_{5}^2)}{2\hbar^3v_F^3} +\frac{7\pi^2T^4}{60\hbar^3v_F^3},\\
P &=& \frac{\epsilon}{3},\\
-en &=& -e\frac{\mu\left(\mu^2+3\mu^2_{5}+\pi^2T^2\right)}{3\pi^2 \hbar^3 v_F^3},\\
-en_{5} &=& -e\frac{\mu_{5}\left(\mu^2_{5}+3\mu^2+\pi^2T^2\right)}{3\pi^2 \hbar^3v_F^3}.
\label{LCHD-general-equilibrium-ee}
\end{eqnarray}
Therefore, the explicit expressions for conductivities in Eqs.~(\ref{model-sigma}) and (\ref{model-sigma5})
read
\begin{eqnarray}
\label{LCHD-general-equilibrium-sigma}
\sigma &=& \frac{3(\mu^2+\mu_5^2)+\pi^2 T^2}{\pi \hbar^2} \tau_{ee},\\
\label{LCHD-general-equilibrium-sigma5}
\sigma_5 &=& \frac{6\mu \mu_5}{\pi \hbar^2}\tau_{ee}.
\end{eqnarray}
It is worth noting that the latter expression is reminiscent to the conductivity of the chiral electric separation effect \cite{Huang:2013iia}.

In an electrically neutral material in equilibrium, the electron charge density $\rho=-en$ should be compensated by the
corresponding background charge density, i.e.,
\begin{equation}
\label{model-compensation-B0}
\rho +\rho_{\rm b} =
-e\frac{\mu\left(\mu^2+3\mu^2_{5}+\pi^2T^2\right)}{3\pi^2 \hbar^3 v_F^3} +\rho_{\rm b}=0.
\end{equation}
When the background magnetic field is nonzero, the Chern--Simons term (\ref{model-CS-charge}) contributes to the electric charge
density. Then, in order to maintain the neutrality of the sample, one should adjust the value of the electric chemical potential.
The corresponding chemical potential $\mu^{(B)}$ is defined by the following relation:
\begin{equation}
\label{model-compensation-B}
\frac{\mu^{(B)}\left[(\mu^{(B)})^2+3\mu^2_{5}+\pi^2T^2\right]}{3\pi^2 \hbar^3 v_F^3}
+\frac{e^2 (\mathbf{b}\cdot\mathbf{B})}{2\pi^2\hbar^2c^2}
-\frac{\mu\left(\mu^2+3\mu^2_{5}+\pi^2T^2\right)}{3\pi^2 \hbar^3 v_F^3}=0.
\end{equation}
Therefore, $\mu^{(B)}$ depends on the magnetic field $\mathbf{B}$, the chiral shift $\mathbf{b}$, the chiral chemical potential
$\mu_5$, temperature $T$, as well as the reference value of the electric chemical potential $\mu$ defined in the absence of the magnetic field. Then, to the linear order in $\mathbf{B}$, the thermodynamic quantities in
Eqs.~(\ref{LCHD-general-equilibrium-be})--(\ref{LCHD-general-equilibrium-ee}) can be obtained simply by replacing $\mu\to\mu^{(B)}$. [Strictly
speaking, the corresponding relations should also receive an explicit dependence on the magnetic field. However, such corrections are
$O(B^2)$ and can be neglected for sufficiently weak fields.]

\begin{figure}[t]
\begin{center}
\includegraphics[width=0.45\textwidth]{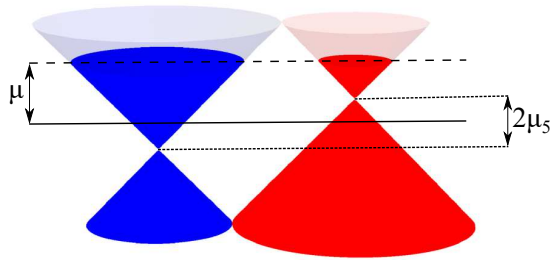}
\caption{The schematic band structure of a Weyl semimetal with broken TR and PI symmetries. In equilibrium, the chiral chemical potential $\mu_5$ is determined by the separation between the Weyl nodes in energy, i.e., $\mu_{5}=eb_0$.}
\label{fig:mu-mu5}
\end{center}
\end{figure}

In the global equilibrium state, the electric current should be absent in Weyl semimetals \cite{Franz:2013}.
By using Eq.~(\ref{model-J-def}) and setting $\mathbf{u}=\mathbf{E}=\bm{\nabla}T=\bm{\nabla}\mu_5=\mathbf{0}$,
this condition gives
\begin{equation}
\label{Minimal-WH-2-J0-compensation}
\mathbf{J}_{\rm eq} = \left(\sigma^{(B)}  - \frac{e^3}{2\pi^2\hbar^2c}b_0 \right)\mathbf{B} = \frac{e^2 \left(\mu_5 -eb_0 \right)\mathbf{B}}{2\pi^2\hbar^2c}= \mathbf{0},
\end{equation}
which is obviously satisfied for $\mu_{5}=eb_0$. Therefore, the chiral chemical potential
in global equilibrium is unambiguously defined in terms of the energy separation between the Weyl nodes.
Such a result is also evident from the band structure of Weyl semimetals, which is schematically shown in Fig.~\ref{fig:mu-mu5}. Indeed, when the Weyl nodes are separated in energy by $2eb_0$, the equilibrium chemical potentials of the individual Weyl nodes are shifted with respect to each other by $2\mu_5$.

%%%%%%%%%%%%%%%%%%
%%%%%%%%%%%%%%%%%%

\subsection{Model setup and boundary conditions}
\label{sec:LCHD-general-setup-and-BC}

In this subsection, we describe the model setup and define the corresponding boundary conditions.
Let us consider an electron fluid flow in a slab of Weyl semimetal infinite in the $x$ and $z$ directions
and of a finite thickness in the $y$ direction, i.e., $0\leq y\leq L_y$.
In this geometry, the local equilibrium variables do not depend on the $x$ and $z$ coordinates when a steady-state regime is maintained.

Further, we specify the boundary conditions (BCs) for the electron fluid velocity and currents.
Obviously, their normal components
should vanish at the surfaces of the slab, i.e.,
\begin{eqnarray}
\label{model-J-BC-y}
J_y(y)\Big|_{y=0,L_y} &=& 0,\\
\label{model-u-BC-y}
u_y(y)\Big|_{y=0,L_y} &=& 0.
\end{eqnarray}
We should also specify the BCs for the components of the
electron fluid velocity parallel to the surfaces.
There are two main types of such BCs \cite{Landau:t6}: (i) the no-slip BCs and (ii) the free-surface (or no-stress)
ones. They are defined by the following equations:
\begin{equation}
\label{model-u-BC-x}
u_x(y)\Big|_{y=0,L_y}=u_z(y)\Big|_{y=0,L_y}=0
\end{equation}
and
\begin{equation}
\label{model-free-BC}
\sum_{k=x,y,z}\left\{\eta \left[\partial_ku_i(y) +\partial_iu_k(y) - \frac{2}{3} \delta_{ik} \left(\bm{\nabla}\cdot\mathbf{u}(y)\right)\right] +\zeta \delta_{ik}\left(\bm{\nabla}\cdot\mathbf{u}(y)\right) -P\delta_{ik} \right\} \hat{n}_k \Bigg|_{y=0,L_y}=0,
\end{equation}
respectively. Here $\hat{\mathbf{n}}$ denotes the surface normal and $i=x,z$. By rewriting Eq.~(\ref{model-free-BC}) in components, one can obtain
\begin{equation}
\label{model-free-BC-x}
\partial_y u_x(y)\Big|_{y=0,L_y}=\partial_y u_z(y)\Big|_{y=0,L_y}=0.
\end{equation}
This relation implies that the tangential forces vanish at the surfaces.
According to the experimental studies performed in WP$_2$ \cite{Gooth:2017}, the most relevant BCs for the electron transport are the
no-slip ones given in Eq.~(\ref{model-u-BC-x}).
Microscopically, this can be understood by the fact that atomically rough edges of the slab could act as impurities allowing for electron
momentum dissipation. However, for the sake of generality, we will also consider the case of the free-surface BCs.

\subsection{Linearized equations}
\label{sec:LCHD-general-eqs}

In this subsection, we present the linearized CHD equations for the model setup defined in the previous subsection. It is very
important that all hydrodynamic and thermodynamic variables in the model at hand depend only on the $y$ coordinate. Then,
by using Maxwell's equations $\bm{\nabla}\times\mathbf{E}(y)=\mathbf{0}$ and $\bm{\nabla}\cdot\mathbf{B}(y)=0$, one can easily check
that $E_x$, $E_z$, and $B_y$ are constant fields. Further, by assuming that the external uniform static electric field is applied in the
$x$ direction and enforcing the boundary conditions for the tangential components of the electric field, we find that $E_z=0$.

It is worthwhile to note that a weak constant external magnetic field does not affect directly the linearized CHD
equations. Indeed, it is included only indirectly via the Chern--Simons term $\propto\mathbf{b}\cdot\mathbf{B}$ in the electric chemical
potential $\mu^{(B)}$ defined by Eq.~(\ref{model-compensation-B}). As a result, its effect in the linear regime is quantitative
rather than qualitative. Therefore, in order to simplify our presentation, the external magnetic field will be ignored in the rest of the paper.
On the other hand, we will calculate the components of the magnetic field $B_x(y)$ and $B_z(y)$ generated by the electric currents in Appendix \ref{sec:app-hydro-vars-MF}. Such a magnetic field could be also used, in principle, to probe the hydrodynamic transport in Weyl semimetals.
(Note that $B_y=0$ for the model setup used.)

Further, we present the nontrivial CHD equations in the linear order in $\mathbf{E}(y)$, $\mathbf{B}(y)$, and $\mathbf{u}(y)$, as well as deviations $\mu(y)-\mu$, $\mu_5(y)-\mu_5$, and $T(y)-T$.
The Navier--Stokes equations for the three components of the flow velocity $\mathbf{u}(y)$ read
\begin{eqnarray}
\label{LCHD-mu5-Euler-x}
&&\eta \partial_{y}^2 u_x(y) -enE_x -\frac{\epsilon+P}{v_F^2\tau} u_x(y) -\frac{\hbar n_5}{4v_F\tau} \partial_y u_z(y)=0,\\
\label{LCHD-mu5-Euler-y}
&&\eta_{y} \partial_{y}^2 u_y(y) -enE_y(y) -\frac{\epsilon+P}{v_F^2\tau} u_y(y) =0,\\
\label{LCHD-mu5-Euler-z}
&&\eta \partial_{y}^2 u_z(y)  -\frac{\epsilon+P}{v_F^2\tau} u_z(y) + \frac{\hbar n_5}{4v_F\tau} \partial_y u_x(y)=0,
\end{eqnarray}
where $\zeta=0$, $\eta_y\equiv4\eta/3$, and $\tau$ is the relaxation time due to intravalley processes that describe the scattering on impurities and/or phonons. Gauss's law is given by
\begin{eqnarray}
\label{LCHD-mu5-Gauss}
&&\varepsilon_e \partial_y E_y(y)=-4\pi\, \left\{e\,\left[n(y)-n\right]  +\frac{e^3}{2\pi^2\hbar^2c^2}
\left[b_x B_x(y)+b_z B_z(y)\right]
\right\}.
\end{eqnarray}
The magnetic field components $B_z(y)$ and $B_x(y)$ are determined by Amper's
law, i.e.,
\begin{eqnarray}
\label{LCHD-mu5-Bz}
&&\partial_y B_z(y) = \mu_{m} \frac{4\pi}{c}\left[ -en u_x(y) +\sigma E_x +\frac{\sigma^{(V)}}{2} \partial_y u_z(y) - \frac{\sigma^{(\epsilon, V)}}{4} \partial_y^2u_x(y) - \frac{e^3}{2\pi^2\hbar^2c}b_z E_y(y) \right],\\
\label{LCHD-mu5-Bx}
&&\partial_y B_x(y) = -\mu_{m} \frac{4\pi}{c}\left\{ -en u_z(y)  -\frac{\sigma^{(V)}}{2} \partial_y u_x(y) -\frac{\sigma^{(\epsilon, V)}}{4} \partial_y^2u_z(y)
+ \frac{e^3}{2\pi^2\hbar^2c} \left[b_x E_y(y)-b_y E_x\right]\right\}.
\end{eqnarray}
Finally, the electric and chiral charge conservation relations are
\begin{eqnarray}
\label{LCHD-mu5-J-continuity}
&&(\bm{\nabla}\cdot\mathbf{J})=-en\partial_yu_y(y) +\sigma \partial_y E_y(y) +\kappa_e \partial_y^2 T(y) + \frac{\sigma_5}{e}\partial_y^2\mu_5(y) =0,\\
\label{LCHD-mu5-J5-continuity}
&&(\bm{\nabla}\cdot\mathbf{J}_5)=-en_5\partial_yu_y(y)+\sigma_5\partial_y E_y(y) +\kappa_{e,5} \partial_y^2 T(y) + \frac{\sigma}{e}\partial_y^2 \mu_5(y) =0,\\
\label{LCHD-mu5-energy}
&&(\epsilon+P)\partial_y u_y = \kappa \partial_y^2 T(y),
\end{eqnarray}
where we used the following linearized
expressions for the currents:
\begin{eqnarray}
\label{LCHD-mu5-J-def}
\mathbf{J} &=& -en\mathbf{u}(y)+\sigma\mathbf{E}(y)+\kappa_e\bm{\nabla} T(y) +\frac{\sigma_5}{e} \bm{\nabla} \mu_5(y) +\sigma^{(V)}\bm{\omega} +\frac{\sigma^{(\epsilon, V)}\left[\bm{\nabla}\times\bm{\omega}\right]}{2} +\frac{e^3\left[\mathbf{b}\times\mathbf{E}(y)\right]}{2\pi^2\hbar^2c},\\
\label{LCHD-mu5-J5-def}
\mathbf{J}_5 &=& -en_{5}\mathbf{u}(y)+\sigma_5\mathbf{E}(y)+\kappa_{e,5}\bm{\nabla} T(y) +\frac{\sigma}{e} \bm{\nabla} \mu_5(y) +\sigma^{(V)}_5\bm{\omega} +\sigma^{(B)}_5\mathbf{B}(y) +\frac{\sigma^{(\epsilon, V)}_5\left[\bm{\nabla}\times\bm{\omega}\right]}{2}.
\end{eqnarray}
Note that, in view of Eq.~(\ref{Minimal-WH-2-J0-compensation}), the CME current $\sigma^{(B)}\mathbf{B}$
is absent in the first equation because it is
compensated by the Chern--Simons term $-e^3b_0\mathbf{B}/(2\pi^2\hbar^2c)$.
Finally, Eq.~(\ref{LCHD-mu5-energy}) corresponds to the linearized version of the energy conservation relation.

In the next two sections, by making use of the linearized CHD equations, we will study the hydrodynamic
electron transport in Weyl semimetals with intact and broken PI symmetries, respectively.

\section{Hydrodynamic flow in the PI symmetric case}
\label{sec:LCHD-hydro-flows-PI}

In this section, we analytically solve the linearized CHD equations defined in Sec.~\ref{sec:LCHD-general}
in Weyl semimetals with preserved PI but broken TR symmetry. This implies that both $b_0$ and $\mu_5$ vanish and
Eqs.~(\ref{LCHD-mu5-Euler-x}), (\ref{LCHD-mu5-Euler-z}), (\ref{LCHD-mu5-Bz})--(\ref{LCHD-mu5-J5-continuity}) could be simplified. In particular, while the $y$ component of the Navier--Stokes equation remains unchanged [see Eq.~(\ref{LCHD-mu5-Euler-y})], the $x$ and $z$ components read
\begin{eqnarray}
\label{LCHD-general-Euler-x}
&&\eta \partial_{y}^2 u_x(y) -enE_x -\frac{\epsilon+P}{v_F^2\tau} u_x(y)
=0,\\
\label{LCHD-general-Euler-z}
&&\eta \partial_{y}^2 u_z(y)  -\frac{\epsilon+P}{v_F^2\tau} u_z(y)=0,
\end{eqnarray}
respectively. The nontrivial components of Ampere's law are
\begin{eqnarray}
\label{LCHD-general-Bz}
&&\partial_y  B_z(y) = \mu_{m} \frac{4\pi}{c}\left[ -en u_x(y) +\sigma E_x -\frac{\sigma^{(\epsilon, V)}}{4} \partial_y^2u_x(y)  - \frac{e^3}{2\pi^2\hbar^2c}b_z E_y(y) \right],\\
\label{LCHD-general-Bx}
&&\partial_y  B_x(y) = -\mu_{m} \frac{4\pi}{c}\left\{ -en u_z(y) -\frac{\sigma^{(\epsilon, V)}}{4} \partial_y^2u_z(y)
+ \frac{e^3}{2\pi^2\hbar^2c} \left[b_x E_y(y)-b_y E_x\right]\right\}.
\end{eqnarray}
The electric and chiral charge conservation relations (\ref{LCHD-mu5-J-continuity}) and (\ref{LCHD-mu5-J5-continuity}), respectively, can be rewritten in the following form:
\begin{eqnarray}
\label{LCHD-general-J-continuity}
&&-en\partial_yu_y(y) +\sigma \partial_y E_y(y) +\kappa_e \partial_y^2 T(y)=0,\\
\label{LCHD-general-J5-continuity}
&&\frac{\sigma}{e}\partial_y^2\mu_5(y) =0.
\end{eqnarray}
Gauss's law and the energy continuity equation
remain unchanged and are given by Eqs.~(\ref{LCHD-mu5-Gauss}) and (\ref{LCHD-mu5-energy}), respectively.
Further, the electric and chiral current densities
(\ref{LCHD-mu5-J-def}) and (\ref{LCHD-mu5-J5-def}) are now given by
\begin{eqnarray}
\label{LCHD-general-J-def}
\mathbf{J} &=& -en\mathbf{u}(y)+\sigma\mathbf{E}(y)+\kappa_e\bm{\nabla} T(y)  +\frac{\sigma^{(\epsilon, V)}\left[\bm{\nabla}\times\bm{\omega}\right]}{2} +\frac{e^3\left[\mathbf{b}\times\mathbf{E}(y)\right]}{2\pi^2\hbar^2c},\\
\label{LCHD-general-J5-def}
\mathbf{J}_5 &=& \frac{\sigma}{e} \bm{\nabla} \mu_5(y) +\sigma^{(V)}_5\bm{\omega} +\sigma^{(B)}_5\mathbf{B}(y).
\end{eqnarray}
The energy conservation relation (\ref{LCHD-mu5-energy}), amended by the appropriate boundary conditions,
defines the spatial distribution of temperature. Note, however, that usually the thermoelectric effects and the
energy conservation relation do not play an important role in the hydrodynamic flow (see, e.g., Ref.~\cite{Lucas:2017idv}).
We will provide below an estimate of their effect for the hydrodynamic transport in Weyl semimetals.

Last but not least, Eq.~(\ref{LCHD-general-J5-continuity}), which is decoupled from the rest of the CHD equations in the PI-symmetric case, defines the spatial distribution of the chiral chemical potential, i.e.,
\begin{equation}
\label{LCHD-general-mu5}
\mu_5(y) = y\frac{\mu_5(L_y)- \mu_5(0)}{L_y} + \mu_5(0),
\end{equation}
where $\mu_5(0)$ and $\mu_5(L_y)$ denote the chiral chemical potential at the surfaces of the slab. It is
reasonable to assume that $\mu_5(y)$ vanishes at the boundaries.
As is obvious from Eq.~(\ref{LCHD-general-mu5}), this leads to $\mu_5(y)=0$.

\subsection{Hydrodynamic flow}
\label{sec:LCHD-Bz0}

In this subsection, we study the hydrodynamic flow in the linearized CHD defined above.
Let us start from the spatial distribution of temperature. Its gradient can be easily
obtained by integrating the energy conservation relation (\ref{LCHD-mu5-energy}), i.e.,
\begin{equation}
\label{LCHD-T-Bz0-Ty}
\partial_y T(y) = \frac{\epsilon+P}{\kappa} u_y(y) +C_T,
\end{equation}
where $C_T=\partial_y T(L_y)$ is determined by the heat flow at the surface. Next, by using the above result and integrating
Eq.~(\ref{LCHD-general-J-continuity}) with the boundary conditions (\ref{model-J-BC-y}) and (\ref{model-u-BC-y}), we obtain
\begin{eqnarray}
\label{LCHD-T-Bz0-Ey}
E_y(y) = \frac{en}{\sigma} u_y(y) -\frac{\kappa_e}{\sigma}\partial_y T(y) - \frac{e^3 E_x b_z}{2\pi^2 \hbar^2 c \sigma}
= E_{y,\rm hydro}(y) +E_{y,\rm thermo}(y)+E_{y,\text{{\tiny CS}}}(y),
\end{eqnarray}
where
\begin{eqnarray}
\label{LCHD-T-Bz0-Ey-hydro}
E_{y,\rm hydro}(y) &=& \left[\frac{en}{\sigma} - \frac{\kappa_e(\epsilon+P)}{\kappa \sigma}\right] u_y(y),\\
\label{LCHD-T-Bz0-Ey-thermo}
E_{y,\rm thermo}(y) &=& -\frac{\kappa_e}{\sigma}C_T,\\
\label{LCHD-T-Bz0-Ey-CS}
E_{y,\text{{\tiny CS}}}(y) &=&- \frac{e^3 E_x b_z}{2\pi^2 \hbar^2 c \sigma}.
\end{eqnarray}
The above equations relate the electric field $E_y(y)$ to the fluid velocity, the temperature gradient, and
the Chern--Simons current (equivalently, the AHE current) determined by the chiral shift.

Having found the electric field $E_y(y)$, we can consider the fluid velocity $\mathbf{u}(y)$ described now by the closed system of
equations (\ref{LCHD-general-Euler-x}), (\ref{LCHD-mu5-Euler-y}), and (\ref{LCHD-general-Euler-z}).
As is easy to show, Eq.~(\ref{LCHD-general-Euler-z}) has only the trivial
solution for the $z$ component of the fluid velocity, i.e., $u_z(y)=0$, when either the free-surface or no-slip BCs are employed.
Owing to the absence of the chiral chemical potential in the PI-symmetric case,
Eq.~(\ref{LCHD-general-Euler-x}) for the $x$ component of the fluid velocity decouples from the rest of the
CHD equations. Its solution reads
\begin{equation}
\label{LCHD-T-Bz0-ux}
u_x(y) = -\frac{v_F^2\tau enE_x}{\epsilon+P} \left[1-\gamma\frac{\cosh{\left(\lambda_x y - \lambda_x L_y/2\right)}}{\cosh{\left(\lambda_x L_y/2\right)}} \right],
\end{equation}
where
\begin{equation}
\label{LCHD-T-Bz0-Euler-x-Sol-lambda}
\lambda_x = \sqrt{\frac{\epsilon+P}{\eta v_F^2 \tau}}
\end{equation}
is an inverse length scale that determines the velocity gradient.
Note that $\gamma=1$ and $\gamma=0$ correspond to the no-slip and free-surface BCs given in
Eqs.~(\ref{model-u-BC-x}) and (\ref{model-free-BC-x}), respectively.
Note that in the former case, the fluid velocity shows a characteristic parabolic-like profile with the maximum
in the middle of the slab and, in the latter case, the velocity is uniform.

It is clear that the effects of the no-slip BCs are localized near the surfaces if $\lambda_xL_y \gtrsim 1$.
On the other hand, if $\lambda_xL_y \ll1$,
the viscous drag effects permeate the whole slab and the
magnitude of the fluid velocity is expected to vary significantly across the slab.
As we will see below, the latter regime could be indeed relevant for the hydrodynamic transport in Weyl semimetals.
The solutions obtained for the different boundary conditions are in agreement with the results in Refs.~\cite{Gurzhi,Gurzhi-effect}
(for similar results in graphene see, e.g., Ref.~\cite{Lucas:2017idv} and references therein).

By substituting the electric field $E_y(y)$ given by Eq.~(\ref{LCHD-T-Bz0-Ey}) into Eq.~(\ref{LCHD-mu5-Euler-y}), we
derive the following result:
\begin{equation}
\label{LCHD-T-Bz0-uy-Sol}
u_y(y) = \frac{en}{\sigma N} \left(\frac{e^3b_zE_x}{2\pi^2\hbar^2c} + \kappa_eC_T \right) \left[1- \frac{\cosh{\left(\lambda_{y} y -\lambda_{y}L_y/2\right)}}{\cosh{\left(\lambda_{y}L_y/2\right)}}\right],
\end{equation}
where
\begin{eqnarray}
\label{LCHD-T-Bz0-tu-def}
N &=& \frac{e^2n^2}{\sigma} + \frac{\epsilon+P}{v_F^2\tau} -\frac{en \kappa_e(\epsilon+P)}{\kappa \sigma},\\
\label{LCHD-T-Bz0-tlambday-def}
\lambda_y &=& \sqrt{\frac{N}{\eta_{y}}}.
\end{eqnarray}
As expected, the obtained \emph{normal fluid flow} quantified by $u_y(y)$ does not depend on the type of the boundary conditions.

The presence of the hydrodynamic flow normal to the slab surfaces is rather unusual. It is instructive, therefore, to clarify  its physical
origin. First, we
note that according to Eq.~(\ref{model-J-BC-y}) the normal component of the electric current density must vanish at the surfaces.
Therefore, in order to compensate the constant Chern--Simons
(or, equivalently, the AHE)
current given
by the last term in Eq.~(\ref{LCHD-general-J-def}), the electric field component normal to the
surface $E_y(y)$ should be generated. Usually, the presence of such a field is enough to compensate the Hall current.
However, in the CHD, the self-consistent solution could be only achieved when
the Navier--Stokes equation (\ref{LCHD-mu5-Euler-y}) and the energy conservation relation (\ref{LCHD-mu5-energy})
are also satisfied. Obviously, Eq.~(\ref{LCHD-mu5-Euler-y}) becomes inhomogeneous at $E_y(y)\neq0$
and allows for a nontrivial normal component of the electron flow velocity $u_y(y)$. The corresponding
result is given by Eq.~(\ref{LCHD-T-Bz0-uy-Sol}) and is determined primarily by the AHE term
proportional to both $b_z$ and $E_x$. [The thermoelectric contribution is described by the term proportional to $\kappa_e$ and vanishes at $C_T=\partial_y T(L_y)=0$.]
Thus, the normal hydrodynamic flow stems
from the self-consistent treatment of the hydrodynamic and electromagnetic sectors of the CHD.

It is instructive to compare the inverse length scale $\lambda_x$ for the longitudinal flow (\ref{LCHD-T-Bz0-Euler-x-Sol-lambda}) and its counterpart for the normal one, i.e., $\lambda_y$.
While the former is determined by the standard hydrodynamic term $(\epsilon+P)/(\eta v_F^2\tau)$, the latter is significantly altered by the electric field $E_y(y)$.
Indeed, as one can see from Eqs.~(\ref{LCHD-T-Bz0-tu-def}) and (\ref{LCHD-T-Bz0-tlambday-def}), the inverse length scale $\lambda_y$ depends on both the conventional hydrodynamic term $(\epsilon+P)/(\eta_yv_F^2\tau)$ and the contributions related to the charged nature of the electron fluid, i.e., $e^2n^2/(\eta_y\sigma)$ and $en \kappa_e(\epsilon+P)/(\eta_y\kappa \sigma)$.
We would like to note that, depending on the model parameters, the hydrodynamic description could be applicable for the longitudinal flow if $\lambda_x^{-1}\gg v_F\tau_{ee}$ but may break down for the normal flow if $\lambda_y^{-1}\lesssim v_F\tau_{ee}$.
In what follows, we will consider only the case when both conditions $\lambda_x^{-1}\gg v_F\tau_{ee}$ and $\lambda_y^{-1}\gg v_F\tau_{ee}$ are satisfied.

By substituting the solutions for $E_y(y)$ as well as the components of the flow velocity into Eqs.~(\ref{LCHD-general-Bz})
and (\ref{LCHD-general-Bx}), it is straightforward to find the generated magnetic fields.
The corresponding expressions are presented in Appendix~\ref{sec:app-hydro-vars-MF}. It is interesting to note that the Chern--Simons current leads
to a magnetic field along the applied electric one.
One could also derive the spatially inhomogeneous part of the electron charge density $n(y)$ by using Gauss's law
(\ref{LCHD-mu5-Gauss}). The resulting expression is given in Appendix~\ref{sec:app-dn}.

Further, by making use of the analytical solutions, let us now investigate the properties and characteristic features of the
hydrodynamic flow in the CHD,
described in terms of the fluid velocity $\mathbf{u}$, the electric current density $\mathbf{J}$, and the electric
potential difference between the surfaces of the slab $U$. In order to obtain numerical estimates,
we use the values of parameters comparable to those in Refs.~\cite{Autes-Soluyanov:2016,Gooth:2017,Kumar-Felser:2017,Razzoli-Felser:2018}, i.e.,
\begin{equation}
\label{LCHD-T-Bz0-realistic-parameters}
v_F\approx1.4\times10^8~\mbox{cm/s}, \quad
b_{\rm latt} \approx 3~\mbox{nm}^{-1}, \quad  b= \frac{\hbar c}{e}\, b_{\rm latt}.
\end{equation}
(Here in order to illustrate the hydrodynamic features, we took $v_F$ an order of magnitude larger than measured for WP$_2$ in Ref.~\cite{Kumar-Felser:2017}.)
The relaxation time $\tau$ and the electric conductivity $\sigma_{\rm exp}$ are, in general, functions of the chemical potentials as well as temperature. According to Ref.~\cite{Gooth:2017}, they range from about $\tau\approx 0.5~\mbox{ns}$ and $\sigma_{\rm exp} \approx 10^{10}~\mbox{S/m}$ at $T=2~\mbox{K}$ to $\tau\approx 5~\mbox{ps}$ and $\sigma_{\rm exp} \approx 2\times10^{8}~\mbox{S/m}$ at $T=30~\mbox{K}$.
As for the dependence on the chemical potentials, in what follows, we will assume that it is weak.
Next, the coefficient $\kappa_0$ describing the violation of the Wiedemann--Franz law in Eq.~(\ref{model-kappa}) ranges from about $\kappa_0\approx0.05$ at $T=4~\mbox{K}$ to $\kappa_0\approx0.45$ at $T=50~\mbox{K}$ \cite{Gooth:2017}.
In addition, we set the electric permittivity $\varepsilon_e\approx13$ (this estimation is based on the dielectric constants of tungsten \cite{Ordal:1988} and phosphorus \cite{Nagahama:1985}) and the magnetic permeability $\mu_m = 1$.

It is worth noting that the CHD equations were formulated and solved in the linear regime in electromagnetic fields.
Therefore, it is important to estimate the characteristic values of the fields that limit the validity of the CHD.
Indeed, in the chiral kinetic theory that was used as a starting point in the derivation of the CHD, the corresponding limitation
can be stated as $|\mathbf{E}|\ll E^{*}$ and $|\mathbf{B}|\ll B^{*}$, where the characteristic electromagnetic
fields are 
\begin{eqnarray}
\label{LCHD-general-E-char-num}
E^{*}&=& \frac{\mu^2}{e\hbar v_F} \approx 1.1\times\left(\frac{\mu}{1~\mbox{meV}}\right)^2~\mbox{kV/m},\\
\label{LCHD-general-B-char-num}
B^{*}&=& \frac{c\mu^2}{e\hbar v_F^2} \approx 0.8\times\left(\frac{\mu}{1~\mbox{meV}}\right)^2~\mbox{mT}.
\end{eqnarray}
When electromagnetic fields (applied externally or induced by the system) become comparable to
these values, the results of the CHD analysis will start to lose their validity.

\begin{figure}[t]
\begin{center}
\includegraphics[width=0.45\textwidth]{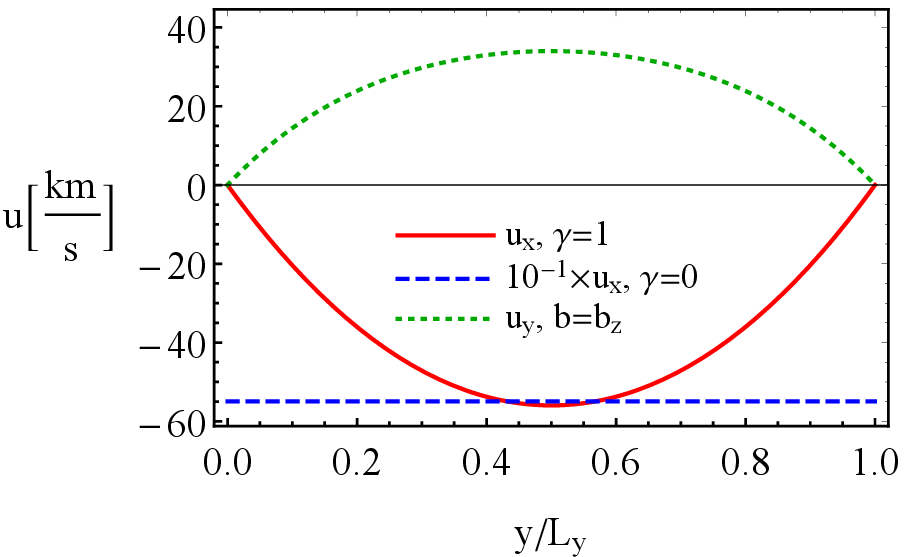}\hfill
\includegraphics[width=0.45\textwidth]{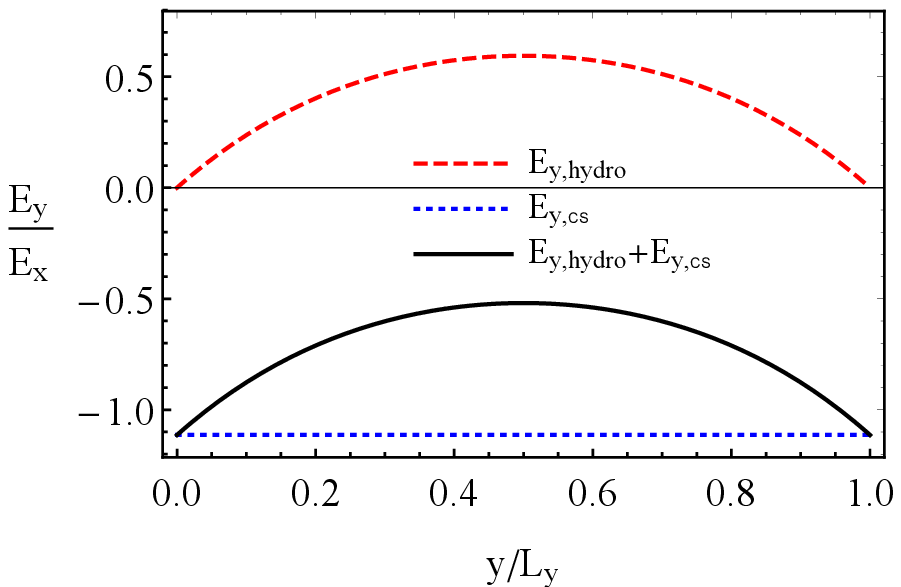}
\caption{Left panel: The dependence of the fluid velocity components $u_x(y)$ and $u_y(y)$ on the spatial coordinate $y$,
assuming either the no-slip BCs ($\gamma=1$) or the free-surface ones ($\gamma=0$).
Right panel: The electric field $E_y(y)$ as a function of $y$.
We also assumed that
$\mathbf{b}\parallel\hat{\mathbf{z}}$ and used the characteristic values of the parameters in
Eq.~(\ref{LCHD-T-Bz0-realistic-parameters}) together with $\mu=10~\mbox{meV}$, $\mu_{5}=0~\mbox{meV}$,
$T=10~\mbox{K}$, $E_x=10~\mbox{V/m}$, $L_y=10~\mu\mbox{m}$, and $C_T=\partial_y T(L_y)=0$.}
\label{fig:LCHD-Bz0-u}
\end{center}
\end{figure}

Next, we present the dependence of the electron fluid velocity $u_x(y)$ and $u_y(y)$ on the $y$ coordinate 
given by
Eqs.~(\ref{LCHD-T-Bz0-ux}) and (\ref{LCHD-T-Bz0-uy-Sol}), respectively, in the left panel of Fig.~\ref{fig:LCHD-Bz0-u}.
As is clear already from Eq.~(\ref{LCHD-T-Bz0-ux}), the chiral shift
does not affect the longitudinal component of the flow velocity $u_x(y)$. This is not surprising
because the $x$ component of the Navier--Stokes equation (\ref{LCHD-general-Euler-x}) decouples
from the rest of the linearized CHD system and describes the usual hydrodynamic flow driven
by the external electric field. Such a flow is unaffected by the topological Chern--Simons terms. As one might expect
in the viscous regime with $\lambda_xL_y \lesssim1$, the profile and the magnitude of $u_x(y)$
depend on the choice of the boundary conditions. The normal component of the flow
$u_y(y)$ is largely driven by the AHE current when there is a nonzero $z$ component
of the chiral shift $b_z$. The velocity $u_y(y)$ across the slab is comparable to $u_x(y)$ and has a similar parabolic-like spatial profile.

Finally, we present the electric field across the slab $E_y(y)$ given by Eq.~(\ref{LCHD-T-Bz0-Ey}) in the right panel of Fig.~\ref{fig:LCHD-Bz0-u}. For $C_T=0$, it consists of the hydrodynamic $E_{y,\rm hydro}(y)$ and purely topological $E_{y,\text{{\tiny CS}}}(y)$ parts.
The interplay of these two components leads to a nontrivial profile where $E_y(y)$ effectively increases near the surfaces of the slab. Obviously, such a spatial profile would be impossible in a nonhydrodynamic regime and provides a hallmark feature of the CHD.
As we will see below, the hydrodynamic part of the electric field will play an important role in the electric current and the electric potential difference between the surfaces of the slab.

%%%%%%%%%%%%%%%%%%
%%%%%%%%%%%%%%%%%%

\subsection{Electric current}
\label{sec:LCHD-T-Bz0}

\begin{figure}[t]
\begin{center}
\includegraphics[width=0.45\textwidth]{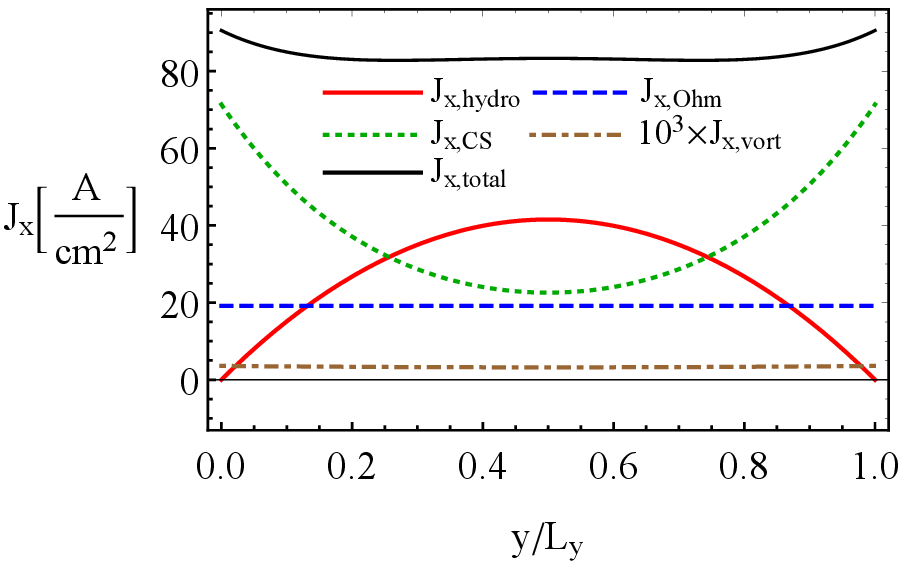}\hfill
\includegraphics[width=0.47\textwidth]{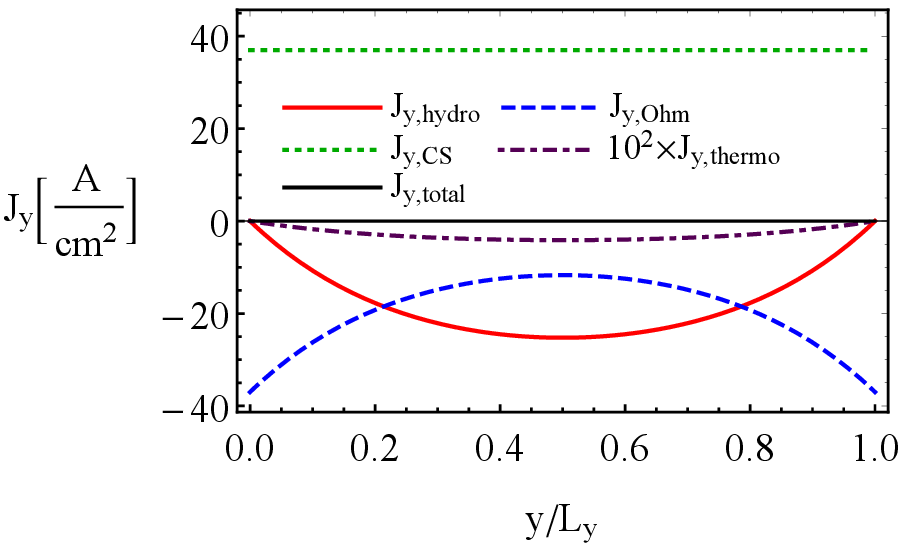}
\caption{The dependence of the electric current density components $J_x(y)$ (left panel) and $J_y(y)$ (right panel) on the spatial coordinate $y$, assuming the
no-slip BCs ($\gamma=1$) and $\mathbf{b}\parallel\hat{\mathbf{z}}$. We used the characteristic
values of the parameters in Eq.~(\ref{LCHD-T-Bz0-realistic-parameters}) together with $\mu=10~\mbox{meV}$,
$\mu_{5}=0~\mbox{meV}$, $T=10~\mbox{K}$, $E_x=10~\mbox{V/m}$, $L_y=10~\mu\mbox{m}$, and $C_T=\partial_y T(L_y)=0$.}
\label{fig:LCHD-Bz0-J-no-slip}
\end{center}
\end{figure}

In this subsection, we discuss the implications of the nontrivial hydrodynamic flow on the electric current.
It can be straightforwardly calculated by making use of the
results for the fluid velocity, the electric field, and the temperature gradient presented in Sec.~\ref{sec:LCHD-Bz0}.
Formally, the total electric current density is given by Eq.~(\ref{LCHD-general-J-def}).
It is convenient to present the total current as
\begin{equation}
\label{LCHD-Bz0-J-def-tot}
\mathbf{J}_{\rm tot} = \mathbf{J}_{\rm hydro} +\mathbf{J}_{\rm Ohm} +\mathbf{J}_{\rm \text{{\tiny CS}}} +\mathbf{J}_{\rm vort}
+\mathbf{J}_{\rm thermo}.
\end{equation}
Here we introduced the shorthand notations for the following qualitatively different types of
contributions: (i) the hydrodynamic current $\mathbf{J}_{\rm hydro}\equiv -en\mathbf{u}(y)$,
(ii) the Ohmic current $\mathbf{J}_{\rm Ohm}\equiv \sigma \mathbf{E}(y)$,
(iii) the topological Chern--Simons current $\mathbf{J}_{\rm \text{{\tiny CS}}}$ that is defined in Eq.~(\ref{model-CS-current}) at $b_0=0$,
(iv) the vortical current $\mathbf{J}_{\rm vort}\equiv \sigma^{(\epsilon,V)} \left[\bm{\nabla}\times \bm{\omega}(y)\right]/2$, and
(v) the thermoelectric current $\mathbf{J}_{\rm thermo}\equiv \kappa_e \bm{\nabla} T(y)$.
For a steady-state solution in the slab geometry the expressions for these currents take the following form:
\begin{eqnarray}
\label{LCHD-Bz0-J-def-hydro}
\mathbf{J}_{\rm hydro} &=& -en\left[u_{x}(y) \hat{\mathbf{x}}+u_{y}(y) \hat{\mathbf{y}} \right],\\
\label{LCHD-Bz0-J-def-Ohm}
\mathbf{J}_{\rm Ohm} &=& \sigma \left[E_{x} \hat{\mathbf{x}}+E_{y}(y) \hat{\mathbf{y}} \right],\\
\label{LCHD-Bz0-J-def-CS}
\mathbf{J}_{\rm \text{{\tiny CS}}} &=& \frac{e^3}{2\pi^2\hbar^2c}\left\{
- b_z E_y(y) \hat{\mathbf{x}}
+ b_z E_x\hat{\mathbf{y}}
+ \left[b_x E_y(y)-b_y E_x\right] \hat{\mathbf{z}}
\right\},\\
\label{LCHD-Bz0-J-def-vort}
\mathbf{J}_{\rm vort} &=& -\frac{\sigma^{(\epsilon,V)}}{4}  \frac{\partial^2 u_{x}(y)}{\partial y^2}\hat{\mathbf{x}},\\
\label{LCHD-Bz0-J-def-thermo}
\mathbf{J}_{\rm thermo} &=& \kappa_e \partial_{y} T(y) \hat{\mathbf{y}},
\end{eqnarray}
where $\hat{\mathbf{x}}$, $\hat{\mathbf{y}}$, and $\hat{\mathbf{z}}$ are the unit vectors in the $x$, $y$, and $z$ directions,
respectively. In the most interesting case of the no-slip BCs ($\gamma=1$), the results for the $x$ and $y$ components
of the electric current densities are shown in Fig.~\ref{fig:LCHD-Bz0-J-no-slip}.

First, let us concentrate on
the longitudinal component of the electric current density, i.e., $J_x$.
As we see from the left panel in
Fig.~\ref{fig:LCHD-Bz0-J-no-slip}, the hydrodynamic and Chern--Simons currents dominate the total current in the $x$ direction.
The Ohmic contribution is few times smaller, albeit it is still comparable to the dominant terms. As for the vortical current
defined in Eq.~(\ref{LCHD-Bz0-J-def-vort}), its contribution to the total current is negligible.
Unlike the other three contributions, however, the Chern--Simons current increases in the vicinity of the surfaces.
Such an effect is directly related to the increase of the electric field $E_y(y)$ caused by the interplay of
the hydrodynamic (\ref{LCHD-T-Bz0-Ey-hydro}) and purely topological (\ref{LCHD-T-Bz0-Ey-CS}) terms (see also the right panel in Fig.~\ref{fig:LCHD-Bz0-u}).
(Note that there is no $x$ component of the thermoelectric current.)

Let us now turn to the normal component of the electric current density, $J_y$. The results for the individual
contributions to $J_y$ are shown in the right panel of Fig.~\ref{fig:LCHD-Bz0-J-no-slip}.
By substituting Eqs.~(\ref{LCHD-T-Bz0-Ty}) and (\ref{LCHD-T-Bz0-Ey}) into the current components (\ref{LCHD-Bz0-J-def-hydro})--(\ref{LCHD-Bz0-J-def-thermo}), one can easily check that
the total current across the slab always vanishes.
However, in the special case where $\mathbf{b}\parallel\hat{\mathbf{z}}$, this is achieved by a compensation of the
hydrodynamic, Ohmic, thermoelectric, and Chern--Simons contributions.
[The normal component of the vortical current (\ref{LCHD-Bz0-J-def-vort}) is absent.]
As expected, the Chern--Simons current is uniform and comparable to the Ohmic and hydrodynamic ones.
This fact is not surprising in view of Eqs.~(\ref{LCHD-T-Bz0-Ey}) and (\ref{LCHD-T-Bz0-uy-Sol}), where the
electric field $E_y(y)$ and the normal flow velocity $u_y(y)$ are also primarily determined by the topological current.
Similarly to $J_{\rm x, \text{{\tiny CS}}}$, the Ohmic current noticeably increases near the surfaces of the slab. However, it is compensated by the corresponding decrease of the other contributions, in particular, the hydrodynamic one. In addition, the thermoelectric current is also nonzero, albeit small.

%%%%%%%%%%%%%%%%%%
%%%%%%%%%%%%%%%%%%

For completeness, let us briefly discuss the case of the free-surface BCs ($\gamma=0$).
As expected, since there is no drag at the boundaries,
the magnitude of the hydrodynamic current becomes larger.
Further, the flow velocity and the currents do not depend on the spatial coordinate $y$. Therefore, it would be difficult to identify the corresponding hydrodynamic features in the longitudinal flow. In either case, the corresponding BCs are unlikely to be realized in real samples of Weyl semimetals. It is worth noting also that the normal flow is not affected by the type of the BCs and is the same for the no-slip and free-surface ones.

As is clear from Eq.~(\ref{LCHD-Bz0-J-def-CS}), the topological Chern--Simons contribution is the only one
that gives rise to the $z$ component of the electric current density. In the presence of the external background electric
field $E_x$, its nature as the AHE current, i.e., $J_{\rm \text{{\tiny CS}},z}\propto b_y E_x$, is obvious at
$b_y\neq0$.
More interestingly, the hydrodynamic flow can strongly alter the Hall current in the $z$ direction, $J_{\rm \text{{\tiny CS}},z}\propto b_xE_y(y)$,
when both $b_x\neq0$ and $b_z\neq0$.
Indeed, according to Eqs.~(\ref{LCHD-T-Bz0-Ey})--(\ref{LCHD-T-Bz0-Ey-CS}),
the normal component of the electric field $E_y(y)$ for $C_T=0$ is determined by the electron flow
velocity $u_y(y)$ and the purely topological AHE current $\propto b_zE_x$.
As seen from Eq.~(\ref{LCHD-T-Bz0-uy-Sol}), the normal flow is also driven by the AHE but is nonuniform across the
slab and its maximum value attained at $y=L_y/2$ depends on the thickness $L_y$.
Therefore, the corresponding electric field component $E_{y,\rm hydro}(y)$ and, consequently, $E_y(y)$ also inherit the nontrivial dependence on $y$.
Clearly, the profile of the corresponding \emph{hydrodynamic AHE} (hAHE) current mimics that
of $E_y(y)$. It is worth noting that, in essence, the hAHE is the anomalous Hall effect in Weyl semimetals modified by the normal flow.

\begin{figure}[t]
\begin{center}
\includegraphics[width=0.45\textwidth]{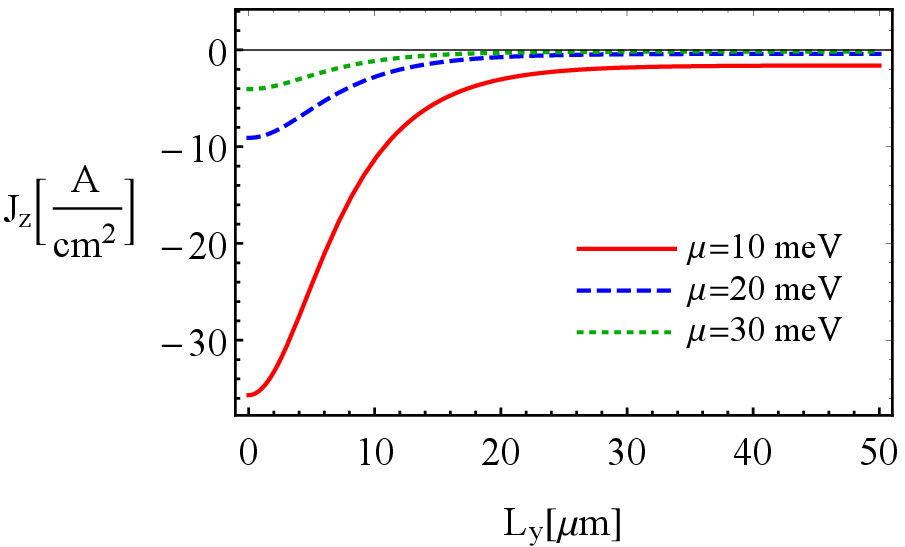}\hfill
\includegraphics[width=0.45\textwidth]{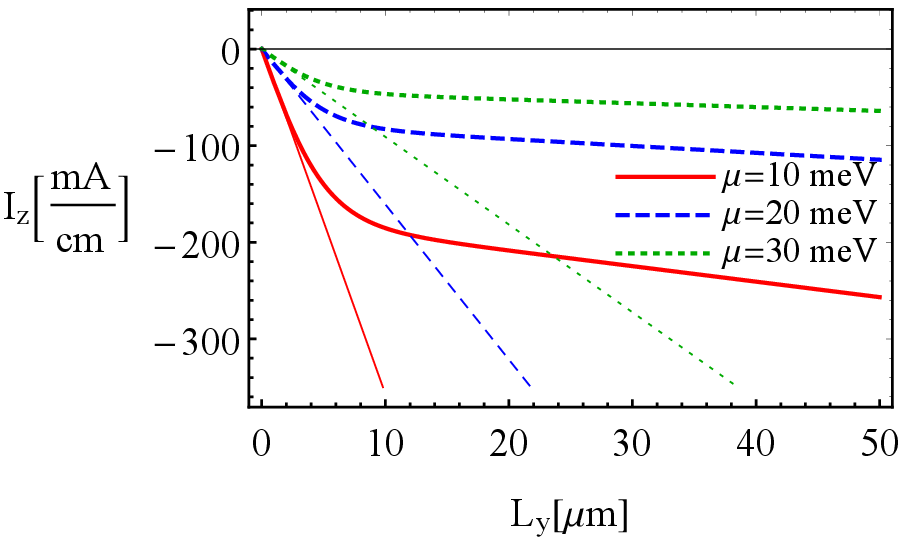}\\
\caption{The dependence of the hydrodynamic AHE current density $J_z(L_y/2)$ (left panel) and the total current per unit length in
the $x$ direction $I_z$ (right panel) on $L_y$.
The red solid lines correspond to $\mu=10~\mbox{meV}$, the blue dashed lines represent the case $\mu=20~\mbox{meV}$, and the green dotted lines show the results for $\mu=30~\mbox{meV}$. The thin lines correspond to the nonhydrodynamic contributions in $I_z$.
We used the parameters in Eq.~(\ref{LCHD-T-Bz0-realistic-parameters}) and set $\mu_{5}=0~\mbox{meV}$, $T=10~\mbox{K}$, $E_x=10~\mbox{V/m}$, $C_T=\partial_y T(L_y)=0$, and $\mathbf{b}=\left\{1,0,1\right\} |\mathbf{b}|/\sqrt{2}$.
}
\label{fig:LCHD-T-Bz0-Jz}
\end{center}
\end{figure}

The dependence of the electric current density in the middle of the slab, $J_z(L_y/2)$, as a function of the thickness $L_y$ is presented in the
left panel of Fig.~\ref{fig:LCHD-T-Bz0-Jz}.
It is also interesting to consider the total electric current per unit length in the $x$ direction, which is obtained by integrating $J_z(y)$ over $y$, i.e.,
\begin{eqnarray}
\label{LCHD-Bz0-Iz-def}
I_z &=& \int_0^{L_y}dy J_z(y) = \frac{e^3 b_x}{2\pi^2 \hbar^2 c \sigma}\left(\frac{e^3b_zE_x}{2\pi^2\hbar^2c} + \kappa_eC_T \right) \left\{ \frac{en}{N\sigma} \left[en - \frac{\kappa_e(\epsilon+P)}{\kappa}\right] \left[L_y- \frac{2}{\lambda_y} \tanh{\left(\frac{\lambda_{y} L_y}{2}\right)} \right]
- L_y\right\}
\nonumber\\
&-&\frac{e^3 b_yE_x}{2\pi^2 \hbar^2 c} L_y.
\end{eqnarray}
The dependence of $I_z$ on the thickness of the slab is shown in the right panel of Fig.~\ref{fig:LCHD-T-Bz0-Jz}.
As we see, both $J_z(L_y/2)$ and $I_z$ depend on the slab thickness and the electric chemical potential.

Let us consider now the dependence of the currents on the chemical potential.
In this connection, it is
instructive to mention that in the Drude regime the corresponding current would be
\begin{equation}
\label{LCHD-Bz0-Iz-Drude-def}
I_z^{(\rm D)} = -\frac{e^3E_x L_y}{2\pi^2 \hbar^2 c} \left(b_y +\frac{e^3 b_xb_z}{2\pi^2 \hbar^2 c \sigma_{\rm Ohm}}\right),
\end{equation}
where $\sigma_{\rm Ohm}$ denotes the conventional Ohmic electric conductivity of Weyl semimetals.
Usually, $\sigma_{\rm Ohm}$ grows with the electric chemical potential decreasing the current in the $z$ direction (see, e.g., Ref.~\cite{Ominato:2014}). A similar behavior is also observed for the hAHE currents shown in Fig.~\ref{fig:LCHD-T-Bz0-Jz}, which decrease with $\mu$.
Therefore, the dependence on the electric chemical potential alone cannot be used to unambiguously confirm the hydrodynamic regime in Weyl semimetals.

However, there is a striking difference between the hydrodynamic and nonhydrodynamic regimes originating from the dependence on the slab thickness $L_y$.
First, we note that the decrease of the electric current density $J_z$ with $L_y$ is notable by itself and is absent in the nonhydrodynamic regime. What is even more important, the experimentally relevant total current per unit length, $I_z$, shows a clear saturation-like behavior. Indeed, as one can see from the right panel in Fig.~\ref{fig:LCHD-T-Bz0-Jz}, the current quickly increases at small $L_y$ and significantly slows down at large $L_y$.
This phenomenon can be explained by the fact that the electric field $E_y(y)$, which drives $J_z$ and $I_z$, is much stronger in the vicinity of the boundaries than in the middle of the slab (see the right panel in Fig.~\ref{fig:LCHD-Bz0-u}).
Therefore, the total current accumulates primarily in the boundary layers leading to the characteristic saturation-like behavior.
Such a nonmonotonic dependence on the slab thickness can be used as a probe of hydrodynamic properties of the Weyl semimetals.

%%%%%%%%%%%%%%%%%%
%%%%%%%%%%%%%%%%%%

It should be noted that the $x$ component of the Chern--Simons current $J_{\rm \text{{\tiny CS}},x}\propto b_zE_y(y)$ (see the left panel in Fig.~\ref{fig:LCHD-Bz0-J-no-slip}) is also nonzero and depends on $L_y$ and $\mu$ in qualitatively the same way as $J_z$.
Unfortunately, there is no clear
method of separating the corresponding hAHE contribution to $J_x$ from the hydrodynamic and Ohmic currents in
the $x$ direction.

Before concluding this subsection, let us mention that, generically, the thermoelectric current has a weak
effect on the fluid velocity. Therefore, a reasonable approximation can be obtained by setting $\kappa_e=0$.
The analytical results in such an approximation are presented in Appendix~\ref{sec:app-hydro-vars-kappae=0}.

\subsection{Hydrodynamic AHE voltage}
\label{sec:LCHD-T-Bz0-pot}

\begin{figure}[t]
\begin{center}
\includegraphics[width=0.45\textwidth]{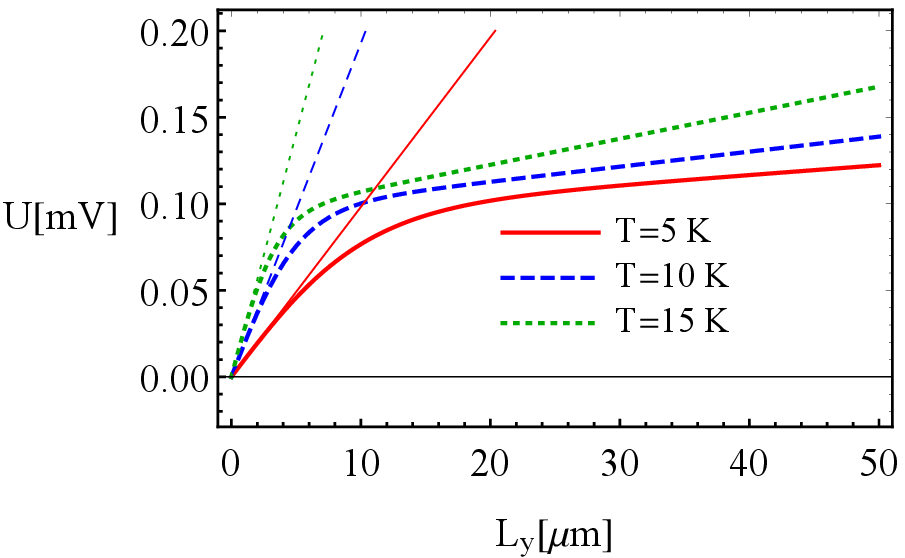}\hfill
\includegraphics[width=0.45\textwidth]{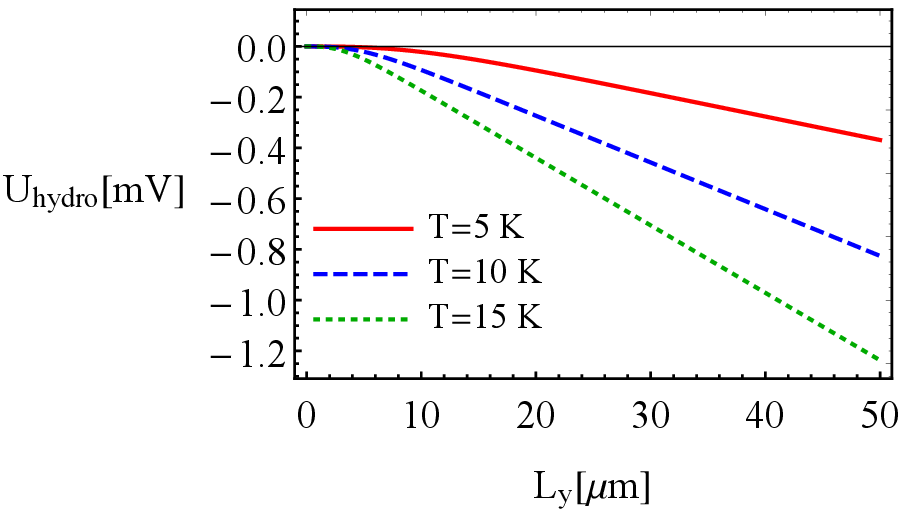}
\caption{The dependence of the hAHE voltage $U$ (left panel) and the hydrodynamic contribution $U_{\rm hydro}$ (right panel) on $L_y$ at various $T$.
The red solid lines correspond to $T=5~\mbox{K}$, the blue dashed lines represent the case $T=10~\mbox{K}$, and the green dotted lines show the results for $T=15~\mbox{K}$.
We used the parameters in Eq.~(\ref{LCHD-T-Bz0-realistic-parameters}) and set $\mu=10~\mbox{meV}$, $\mu_{5}=0$, $E_x=10~\mbox{V/m}$, $C_T=\partial_y T(L_y)=0$, and $\mathbf{b}\parallel\hat{\mathbf{z}}$.}
\label{fig:LCHD-T-Bz0-dphi}
\end{center}
\end{figure}

Similarly to the usual Hall effect, its anomalous counterpart induces a nonzero electric potential difference between the opposite surfaces of the slab.
The explicit expression for such a hAHE voltage can be obtained
by calculating the line integral of the electric field across the slab
\begin{eqnarray}
\label{LCHD-Bz0-Deltaphi-def}
U = -\int_0^{L_y}dy E_y(y) = U_{\rm hydro}+U_{\text{{\tiny CS}}},
\end{eqnarray}
where, for convenience, we separated the normal flow contribution from the purely topological (nonhydrodynamic) one, i.e.,
\begin{eqnarray}
\label{LCHD-Bz0-Deltaphi-hydro}
U_{\rm hydro} &=& -\left(\frac{e^3b_zE_x}{2\pi^2\hbar^2c} + \kappa_eC_T \right) \frac{en}{N\sigma^2} \left[en - \frac{\kappa_e(\epsilon+P)}{\kappa}\right] \left[L_y- \frac{2}{\lambda_y} \tanh{\left(\frac{\lambda_{y} L_y}{2}\right)}
\right],\\
\label{LCHD-Bz0-Deltaphi-CS}
U_{\text{{\tiny CS}}}  &=& \frac{L_y}{\sigma}\left(\frac{e^3b_zE_x}{2\pi^2\hbar^2c} + \kappa_eC_T \right).
\end{eqnarray}
We present the hAHE voltage $U$ as a function of $L_y$ in the left panel of Fig.~\ref{fig:LCHD-T-Bz0-dphi} for several values
of $T$. As one can see, the dependence of $U$ on the slab thickness is nonmonotonic and shows the same saturation-like behavior as the total electric current $I_z$ (cf. Figs.~\ref{fig:LCHD-T-Bz0-Jz} and \ref{fig:LCHD-T-Bz0-dphi}). Obviously, it also originates from the inhomogeneous profile of the electric field $E_y(y)$ that is enhanced near the surfaces of the slab (see the right panel in Fig.~\ref{fig:LCHD-Bz0-u}). In order to better understand this phenomenon, we also present the hydrodynamic $U_{\rm hydro}$ and nonhydrodynamic $U_{\text{{\tiny CS}}}$ contributions; see the right panel and the thin lines in the left panel of Fig.~\ref{fig:LCHD-T-Bz0-dphi}, respectively.
As expected, the nonhydrodynamic voltage $U_{\text{{\tiny CS}}}$ linearly increases with the slab thickness.
On the other hand, the hydrodynamic part $U_{\rm hydro}$ tends to compensate the nonhydrodynamic voltage.
Indeed, by neglecting the subleading term $(\epsilon+P)/(v_F^2\tau)$ in the coefficient $N$ given by Eq.~(\ref{LCHD-T-Bz0-tu-def}), one can check that the linear in $L_y$ term in $U_{\rm hydro}$ [i.e., the first term in the last square brackets in Eq.~(\ref{LCHD-Bz0-Deltaphi-hydro})] cancels $U_{\text{{\tiny CS}}}$.
Therefore, the steplike profile is determined primarily by the second term in the last square brackets in Eq.~(\ref{LCHD-Bz0-Deltaphi-hydro}) and quickly reaches the constant value $\propto2/\lambda_y$.
In general, however, such a cancellation is not exact, which explains the slow linear increase of the voltage at large $L_y$.
In addition, we note that the profile of the hAHE voltage becomes less pronounced at high temperatures.
Such an effect is due to the fact that the term $(\epsilon+P)/(v_F^2\tau)$ in the normal fluid velocity starts to dominate.
Still, there is a clear difference between the hAHE at small and large thicknesses of the slab.
Thus, similarly to the total electric current $I_z$, the saturation-like behavior of the hAHE voltage can be used to investigate the hydrodynamic features of the electron transport in Weyl semimetals.

%%%%%%%%%%%%%%%%%%
%%%%%%%%%%%%%%%%%%

\section{Hydrodynamic flow in Weyl semimetals with broken PI symmetry}
\label{sec:LCHD-hydro-flows-PI-broken}

In this section, we study the hydrodynamic flow of the chiral electrons in Weyl semimetals with broken PI and TR symmetries.
In this case $b_0\neq0$ and, as follows from Eq.~(\ref{Minimal-WH-2-J0-compensation}), the chiral chemical potential
$\mu_{5}=eb_0$ is nonzero. The corresponding system of the CHD equations is given by Eqs.~(\ref{LCHD-mu5-Euler-x})--(\ref{LCHD-mu5-energy}).

Let us start by reexpressing the electric and chiral current continuity relations (\ref{LCHD-mu5-J-continuity}) and
(\ref{LCHD-mu5-J5-continuity}) in a more convenient form. Multiplying Eq.~(\ref{LCHD-mu5-J5-continuity}) by $\sigma_5/\sigma$ and subtracting
it from Eq.~(\ref{LCHD-mu5-J-continuity}), we obtain the relation that does not contain derivatives of the chiral chemical potential, i.e.,
\begin{equation}
\label{LCHD-mu5-2-J-continuity-3}
-e\tilde{n}\partial_yu_y(y) +\tilde{\kappa}_e \partial_y^2 T(y) +\tilde{\sigma} \partial_y E_y(y)=\left[-e\tilde{n} +\frac{\tilde{\kappa}_e(\epsilon+P)}{\kappa} \right] \partial_yu_y(y) +\tilde{\sigma} \partial_y E_y(y) =0.
\end{equation}
Here we used Eq.~(\ref{LCHD-mu5-energy}) and introduced the following shorthand notations:
\begin{eqnarray}
\label{LCHD-mu5-2-J-continuity-tn}
\tilde{n} &=& n -n_5\frac{\sigma_5}{\sigma},\\
\label{LCHD-mu5-2-J-continuity-tsigma}
\tilde{\sigma} &=& \sigma -\frac{\sigma_5^2}{\sigma},\\
\label{LCHD-mu5-2-J-continuity-tkappae}
\tilde{\kappa}_e &=& \kappa_e - \frac{\kappa_{e,5}\sigma_5}{\sigma}.
\end{eqnarray}
In essence, Eq.~(\ref{LCHD-mu5-2-J-continuity-3}) is a modified electric charge conservation relation and is analogous to Eq.~(\ref{LCHD-general-J-continuity}) in the PI-symmetric case. By making use of such a similarity, we find that the solution for $E_y(y)$ is given by the same expression
as in Eq.~(\ref{LCHD-T-Bz0-Ey}), but with the following replacements: $n\to\tilde{n}$, $\sigma\to\tilde{\sigma}$,
and $\kappa_e\to\tilde{\kappa}_e$.
The fluid velocity $u_y(y)$ is
\begin{equation}
\label{LCHD-T-Bz0-uy-Sol-PI}
u_y(y) = \frac{en}{\tilde{\sigma} \tilde{N}} \left(\frac{e^3b_zE_x}{2\pi^2\hbar^2c} + \tilde{\kappa}_eC_T \right) \left[1- \frac{\cosh{\left(\tilde{\lambda}_{y} y -\tilde{\lambda}_{y}L_y/2\right)}}{\cosh{\left(\tilde{\lambda}_{y}L_y/2\right)}}\right],
\end{equation}
where the coefficients $\tilde{N}$ and $\tilde{\lambda}_y$ are given by
\begin{eqnarray}
\label{LCHD-T-Bz0-ttN-def}
\tilde{N} &=& \frac{e^2n \tilde{n}}{\tilde{\sigma}} + \frac{\epsilon+P}{v_F^2\tau} -\frac{en \tilde{\kappa}_e (\epsilon+P)}{\kappa \tilde{\sigma}} ,\\
\label{LCHD-T-Bz0-ttlambday-def}
\tilde{\lambda}_y &=& \sqrt{\frac{\tilde{N}}{\eta_{y}}}.
\end{eqnarray}

Further, we consider the components of the electron flow velocity parallel to the slab surfaces, i.e., $u_x(y)$ and $u_z(y)$, described by Eqs.~(\ref{LCHD-mu5-Euler-x}) and (\ref{LCHD-mu5-Euler-z}), respectively. By introducing a complex variable $u_{xz}(y)=u_x(y)+i\,u_z(y)$, we obtain the following equation:
\begin{equation}
\label{LCHD-mu5-2-Euler-xz}
\eta \partial_{y}^2 u_{xz}(y) -enE_x -\frac{\epsilon+P}{v_F^2\tau} u_{xz}(y) +i\frac{\hbar n_5}{4v_F\tau} \partial_y u_{xz}(y)=0.
\end{equation}
Its solution reads
\begin{equation}
\label{LCHD-mu5-2-Euler-xz-sol}
u_{xz}(y) = -\frac{v_F^2\tau e n E_x}{\epsilon+P} \left(1 - \gamma e^{\lambda_{+}y}\frac{1-e^{\lambda_{-}L_y}}{e^{\lambda_{+}L_y}-e^{\lambda_{-}L_y}} +\gamma e^{\lambda_{-}y} \frac{1-e^{\lambda_{+}L_y}}{e^{\lambda_{+}L_y}-e^{\lambda_{-}L_y}}\right),
\end{equation}
where
\begin{equation}
\label{LCHD-mu5-2-Euler-lambdaxz}
\lambda_{\pm} = -\frac{i \hbar n_5}{8 v_F \tau \eta} \pm \frac{1}{2}\sqrt{\frac{4(\epsilon+P)}{v_F^2\tau \eta} - \frac{\hbar^2n_5^2}{16v_F^2\tau^2\eta^2}}.
\end{equation}
Obviously, the $x$ and $y$ components of the fluid velocity can be found by separating the real and imaginary parts in
Eq.~(\ref{LCHD-mu5-2-Euler-xz-sol}), i.e., $u_x(y)=\mbox{Re}\left[u_{xz}(y)\right]$ and
$u_z(y)=\mbox{Im}\left[u_{xz}(y)\right]$.
It is worth noting that neither $u_x(y)$ nor $u_z(y)$ are affected by
the chiral shift $\mathbf{b}$.

For the experimentally relevant parameters, the second term under the square root in Eq.~(\ref{LCHD-mu5-2-Euler-lambdaxz}) is negligible compared to the first one. Therefore, the real and imaginary parts of $u_{xz}(y)$ can be easily separated.
By making use of the notations
\begin{eqnarray}
\label{LCHD-mu5-2-Euler-lambdaxz-Re}
\lambda_{\rm R}&=&\mbox{Re}\left(\lambda_{+}\right)= \frac{1}{2}\sqrt{\frac{4(\epsilon+P)}{v_F^2\tau \eta} - \frac{\hbar^2n_5^2}{16v_F^2\tau^2\eta^2}},\\
\label{LCHD-mu5-2-Euler-lambdaxz-Im}
\lambda_{\rm I}&=&\mbox{Im}\left(\lambda_{+}\right)= -\frac{\hbar n_5}{8v_F \tau \eta},
\end{eqnarray}
we derive the following expressions for the individual components of the flow velocity:
\begin{eqnarray}
\label{LCHD-mu5-2-Euler-ux}
u_x(y) &=& \frac{en\tau v_F^2 E_x}{\epsilon+P} \left[1-\coth{\left(\lambda_{\rm R} L_y\right)}\right] e^{L_y \lambda_R} \Bigg\{ \sinh{\left(L_y\lambda_R\right)}
\left[1-\gamma e^{(L_y-y)\lambda_{\rm R}} \cos{\left((L_y-y)\lambda_{\rm I}\right)}\right] \nonumber\\
&-&\gamma \sinh{\left((L_y-y)\lambda_R\right)} \cos{\left(y \lambda_{\rm I}\right)}  \Bigg\},\\
\label{LCHD-mu5-2-Euler-uz}
u_z(y) &=& \gamma \frac{en\tau v_F^2 E_x}{\epsilon+P} \left[1-\coth{\left(\lambda_{\rm R} L_y\right)}\right] e^{L_y \lambda_R}
\Bigg\{ \sinh{\left(y\lambda_R\right)} \sin{\left((L_y-y)\lambda_{\rm I}\right)} \nonumber\\
&-& \sinh{\left((L_y-y)\lambda_R\right)} \sin{\left(y \lambda_{\rm I}\right)}  \Bigg\}.
\end{eqnarray}
As is easy check, the profile for $u_x(y)$ in Eq.~(\ref{LCHD-T-Bz0-ux}) is reproduced when $n_5=0$
(i.e., $\lambda_{\rm I}=0$). We also note that the $z$ component of the fluid velocity in Eq.~(\ref{LCHD-mu5-2-Euler-uz})
vanishes when either $\mu_5=0$ or the free-surface BCs ($\gamma=0$) are used.

\begin{figure}[t]
\begin{center}
\includegraphics[width=0.45\textwidth]{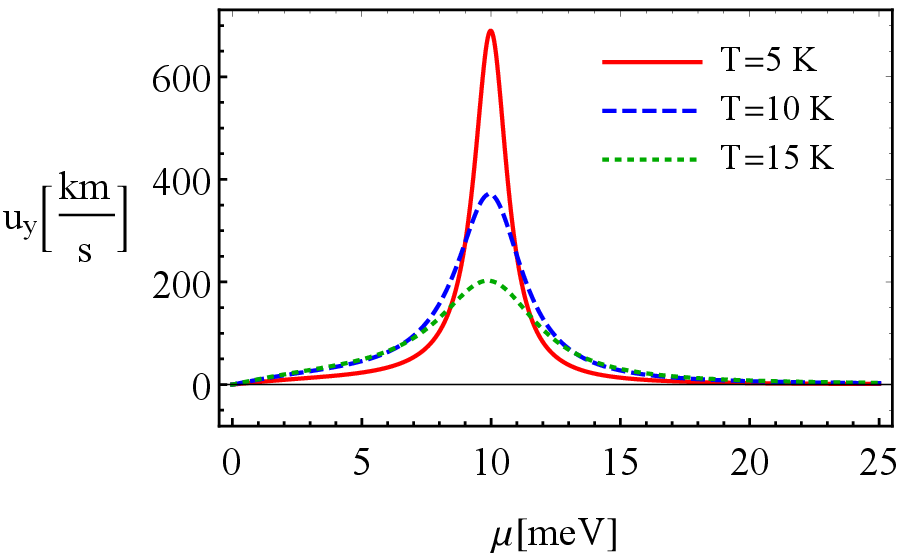}\hfill
\includegraphics[width=0.45\textwidth]{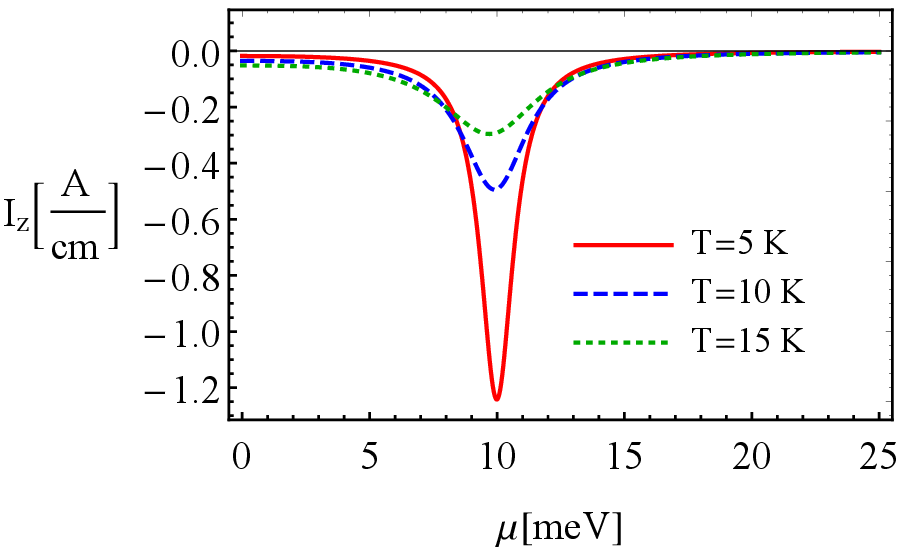}
\caption{The dependence of the normal component of the fluid velocity $u_y(L_y/2)$ and the total electric current per unit length in
the $x$ direction $I_z$ on $\mu$ for several values of $T$.
The red solid lines correspond to $T=5~\mbox{K}$, the blue dashed lines represent the case $T=10~\mbox{K}$, and the green dotted lines show the results at $T=15~\mbox{K}$.
We used the parameters in Eq.~(\ref{LCHD-T-Bz0-realistic-parameters}) and set $\gamma=1$, $\mu_5=10~\mbox{meV}$, $E_x=10~\mbox{V/m}$, $L_y=10~\mu\mbox{m}$, and $C_T=\partial_y T(L_y)=0$.}
\label{fig:LCHD-mu5-2-u-from-mu5}
\end{center}
\end{figure}

It is interesting to consider the dependence on $\mu$ of the mid-stream velocity component $u_y(L_y/2)$ and the total current per unit length in
the $x$ direction $I_z$. Such a dependence for a nonzero chiral chemical potential $\mu_5$ is presented in the left and right panels of Fig.~\ref{fig:LCHD-mu5-2-u-from-mu5}, respectively.
The normal flow velocity $u_y(L_y/2)$ is nonmonotonic with a well-pronounced peak around
$\mu\approx\mu_5$. This striking feature is related to the fact that the effective conductivity
(\ref{LCHD-mu5-2-J-continuity-tsigma}) becomes small at $\mu=\mu_5$, i.e.,
\begin{equation}
\label{LCHD-mu5-2-tsigma-T0}
\lim_{\mu\to\mu_5}\tilde{\sigma} \approx \frac{2\pi T}{\hbar} - \frac{\pi^3 T^3}{6\hbar \mu_5^2} + O\left(\frac{T}{\mu_5}\right)^5
\end{equation}
and vanishes for $T\to0$. The nontrivial normal flow velocity $u_y(L_y/2)$ leads to a similar peak in the total electric current $I_z$ (see
the right panel in Fig.~\ref{fig:LCHD-mu5-2-u-from-mu5}). Such a feature again signifies the importance of the hydrodynamic flow for the
electric current and is another result of the CHD. We note, however, that this peak relies on our model for the intrinsic conductivities $\sigma$ and $\sigma_5$ given in Eqs.~(\ref{model-sigma}) and (\ref{model-sigma5}), respectively.
Therefore, it would be very interesting to investigate whether such a phenomenon
persists in real materials.
In addition, we found that, in view of Eq.~(\ref{LCHD-mu5-2-Euler-uz}), the hAHE current also obtains the hydrodynamic contribution $-enu_z(y)$ at $\mu_5\neq0$.
However, the effect of the corresponding velocity $u_z(y)$ is negligible at the used parameters.

%%%%%%%%%%%%%%%%%%
%%%%%%%%%%%%%%%%%%

In passing, we discuss the chiral chemical potential $\mu_5(y)$. By using the energy conservation relation (\ref{LCHD-mu5-energy}) and Eq.~(\ref{LCHD-mu5-2-J-continuity-3}),
the chiral current continuity relation (\ref{LCHD-mu5-J5-continuity}) can be rewritten as
\begin{eqnarray}
\label{LCHD-mu5-J5-continuity-3}
&&\partial_y^2\mu_5(y) = -\frac{e}{\sigma}\left\{-en_5 +\frac{\kappa_{e,5}(\epsilon+P)}{\kappa} +\frac{\sigma_5}{\tilde{\sigma}} \left[e\tilde{n} -\frac{\tilde{\kappa}_e (\epsilon+P)}{\kappa}\right] \right\} \partial_yu_y(y).
\end{eqnarray}
This equation, amended by the corresponding boundary conditions [see the discussion below Eq.~(\ref{LCHD-general-mu5})], defines the spatial
profile of the chiral chemical potential. Therefore, the normal flow could also lead to the deviations of $\mu_5(y)$ from its global equilibrium value $eb_0$.

\section{Summary and discussions}
\label{sec:Summary}

In this paper, we studied a steady-state flow of chiral electrons in a Weyl semimetal slab by using
the consistent hydrodynamic theory proposed in Ref.~\cite{Gorbar:2017vph}.
The framework contains the viscous, anomalous, and vortical effects, as well as the intrinsic conductivities in the electric and chiral current densities.
The distinctive feature of the consistent hydrodynamics is, however, the inclusion of the topological Chern--Simons contributions that
introduces the dependence on the energy $2b_0$ and momentum $2\mathbf{b}$ separations between
the Weyl nodes. Such a dependence is absent in the Navier--Stokes equation and the energy conservation
relation. Instead, the topological contributions influence
the hydrodynamics only indirectly via Maxwell's equations.

In a finite-thickness slab with the no-slip boundary conditions, we found that the electron fluid velocity $\mathbf{u}$ has a
characteristic parabolic-like spatial profile when the external electric field $\mathbf{E}$ is applied
in the direction parallel to the slab surfaces.
On the other hand, as expected for the free-surface boundary conditions, the flow velocity stays uniform
and, if the thickness of the slab $L_y$ is small, its magnitude is significantly larger than in the case of
the no-slip boundary conditions.

Most interestingly, we revealed that the Chern--Simons current in Weyl semimetals can qualitatively affect the hydrodynamic flow.
Remarkably, it leads to the appearance of the \emph{normal flow} velocity (i.e., directed perpendicularly to the slab surfaces) inside the slab.
In essence, such a hydrodynamic flow originates from the anomalous Hall effect
that is driven by the external electric field and the chiral shift when the latter has a nonzero component
parallel to the slab surfaces and perpendicular to $\mathbf{E}$.
Since the normal component of the electric current density vanishes at the surfaces, the normal component of the anomalous Hall current should be always compensated in a steady state.
Usually, this is achieved by the generation of the corresponding electric field that leads to the Ohmic current.
However, in the consistent hydrodynamic framework used in this paper, the self-consistent treatment necessary leads to the formation of the electron flow normal to the surface.
Such a flow has a parabolic-like spatial profile and vanishes at the slab surfaces.
Further, the corresponding velocity decreases with the electric chemical potential $\mu$ and shows a nonmonotonic dependence on $L_y$.
In particular, it quickly grows at small values of the slab thickness and slows down at sufficiently large ones.
Although the normal flow is remarkable by itself, it might be difficult to observe directly in experiments.
Therefore, we proposed two other potentially observable effects related to the hydrodynamic transport of chiral electrons in Weyl materials.

The first effect
is the generation of an electric current in the direction parallel to the slab surfaces but perpendicular to $\mathbf{E}$.
In essence, such a current is an anomalous Hall one, but strongly modified by the normal flow when the chiral shift has nonzero components
in the two directions parallel to the slab surfaces.
Therefore, we called this phenomenon a \emph{hydrodynamic AHE} (hAHE).
The hydrodynamic nature of the corresponding current is evident from its dependence on the slab thickness. In particular, at sufficiently low temperatures, the total hAHE current per unit length in the $x$ direction
shows a characteristic steplike profile as a function of $L_y$.
What is also important, is that it is not obscured by the standard Ohmic current and should be easily observed by applying additional electric contacts to the slab that is large but finite in the $z$ direction.
The hAHE current is strongly enhanced in Weyl semimetals with a broken parity-inversion symmetry when the electric chemical potential
matches the separation of the Weyl nodes in energy.
In addition, the current generates a magnetic field directed along the external electric one, which is also sensitive to the hydrodynamic flow.

In the slab geometry studied here, the nontrivial topological properties of the hydrodynamic transport in the Weyl semimetals can be
also revealed by measuring the Hall voltage $U$ between the surfaces of the slab. While such a voltage originates from the anomalous Hall
effect, we found that $U$ is also strongly affected by the hydrodynamic flow of electrons.
Similarly to the total current per unit length in the $x$ direction, this hAHE voltage quickly increases at small values of the slab thickness but saturates at the large ones.
Such a behavior with two different slopes is allowed by the normal flow and is absent in the conventional Drude regime.
We argue that, in principle, the nonlinear dependence of the hAHE voltage on $L_y$ can be used to experimentally probe and confirm the hydrodynamic regime.

It is worth emphasizing that in the present study we used the consistent hydrodynamic description
of the left- and right-handed electrons as parts of a single fluid with a common flow velocity
$\mathbf{u}$. In general, the presence of the chirality-dependent forces might lead to a breakdown of such
an approximation. Then, a two-fluid description with velocities $\mathbf{u}$ and $\mathbf{u}_5$ might be needed.
For example, the corresponding effects might become important in the presence of strain-induced pseudomagnetic
fields, which exert chirality-dependent Lorentz forces on quasiparticles. In principle, the separation of chiral fluids could also be driven by the vorticity-related terms in the presence of an external magnetic field, although the splitting is expected to be rather small.
The investigation of the corresponding hydrodynamic effects within a two-fluid model will be reported elsewhere.

Before concluding, let us briefly discuss the role of boundaries in the description of the hydrodynamic
flow. While we employed the conventional no-slip and free-surface boundary conditions,
one might question whether they indeed are realized in Weyl semimetals.
In particular,
the situation might be more complicated because Weyl semimetals host topologically nontrivial Fermi
arc surface states \cite{Savrasov} (for recent reviews, see
Refs.~\cite{Yan-Felser:2017-Rev,Hasan-Huang:2017-Rev,Armitage-Vishwanath:2017-Rev}). Such states could, in principle, influence
the hydrodynamic flow in the vicinity of the slab surfaces and modify the boundary conditions.
The flow of electrons near the slab surfaces could be also affected by the pseudomagnetic
field $\mathbf{B}_5$ that is expected near the boundary of the sample because of an
abrupt change in the chiral shift \cite{Grushin-Vishwanath:2016}. The corresponding surface currents
could modify our predictions for the electric potential, the generated magnetic field, and the electron charge density near the surfaces of the slab. It would be interesting, therefore, to
address rigorously the problem of the boundary conditions and the effect of the surface
states on the hydrodynamics of chiral electrons in Weyl semimetals.
The corresponding study, however, is beyond the scope of this paper.

\begin{acknowledgments}
The work of E.V.G. was partially supported by the Program of Fundamental Research of the
Physics and Astronomy Division of the National Academy of Sciences of Ukraine.
The work of V.A.M. and P.O.S. was supported by the Natural Sciences and Engineering Research Council of Canada.
The work of I.A.S. was supported by the U.S. National Science Foundation under Grants PHY-1404232
and PHY-1713950.
\end{acknowledgments}

\appendix

\section{Equations of the steady-state consistent hydrodynamics}
\label{sec:app-CHD-eqs}

In this appendix, we present the steady-state equations of the consistent hydrodynamics (CHD) amended
by the viscosity effects, including the electric and chiral charge conservation relations, the Navier--Stokes equation,
and the energy conservation relation.

The electric and chiral charge conservation relations are given by
\begin{eqnarray}
\label{model-J-conserv-eq}
\left(\bm{\nabla}\cdot\mathbf{J}\right) &=& 0, \\
\label{model-J5-conserv-eq}
\left(\bm{\nabla}\cdot\mathbf{J}_5\right) &=& -\frac{e^3 (\mathbf{E}\cdot\mathbf{B})}{2\pi^2 \hbar^2 c},
\end{eqnarray}
where the second equation is related to the celebrated chiral anomaly~\cite{Adler,Bell-Jackiw} and describes the chiral charge inflow in the presence of electric $\mathbf{E}$ and magnetic $\mathbf{B}$ fields. We note that
$\mathbf{J}$ and $\mathbf{J}_5$ are the \emph{total} electric and chiral current densities, respectively.
Their explicit form is given in the main text; see Eqs.~(\ref{model-J-def}) and (\ref{model-J5-def}).
By definition, $e$ is the absolute value of the electron charge and $c$ is the speed of light.

The Navier--Stokes equation, which is nothing else but the Euler equation \cite{Gorbar:2017vph} amended
by the viscous terms, is given by
\begin{eqnarray}
\label{model-Euler}
&&-\eta \Delta \mathbf{u} -\left(\zeta+\frac{\eta}{3}\right) \bm{\nabla}\left(\bm{\nabla}\cdot\mathbf{u}\right)+\bm{\nabla}P + \frac{4}{15v_F} \left[\sum_{j=1}^3 B_j \bm{\nabla} u_j + (\mathbf{B}\cdot\bm{\nabla})\mathbf{u} +\mathbf{B} (\bm{\nabla}\cdot \mathbf{u})\right] \sigma^{(\epsilon, B)} \nonumber\\
&&+ \frac{c \left[\bm{\nabla}\times\mathbf{E}\right] \sigma^{(\epsilon, B)}}{3v_F}
+\frac{2}{3v_F}\sum_{j=1}^3 u_j\bm{\nabla}B_j \sigma^{(\epsilon, B)}
-\frac{4\sigma^{(\epsilon, B)}}{15v_F} \left[\sum_{j=1}^3 u_j\bm{\nabla}B_j +(\mathbf{u}\cdot\bm{\nabla})\mathbf{B}\right]
+\frac{5\sigma^{(\epsilon, u)}\bm{\nabla}B^2}{2} \nonumber\\
&&+ \left[(\mathbf{B}\cdot\bm{\nabla})\bm{\omega} + \sum_{j=1}^3 B_j\bm{\nabla} \omega_j  \right] \frac{\sigma^{(\epsilon, V)}}{5c} -\sum_{j=1}^3 \frac{\sigma^{(\epsilon, V)}\omega_j \bm{\nabla}B_j}{2c}
-\frac{\sigma^{(\epsilon, V)}}{5c} \left[\sum_{j=1}^3\omega_j\bm{\nabla}B_j + (\bm{\omega}\cdot\bm{\nabla})\mathbf{B} \right] \nonumber\\
&& =-en\mathbf{E} +\frac{1}{c}\left[\mathbf{B}\times \left(en\mathbf{u}
-\frac{\sigma^{(V)}\bm{\omega}}{3}
\right)\right] +\frac{\sigma^{(B)} \mathbf{u} (\mathbf{E}\cdot\mathbf{B})}{3v_F^2} +\frac{5c \sigma^{(\epsilon, u)} (\mathbf{E}\cdot\mathbf{B})\bm{\omega}}{v_F}
 -\frac{(\epsilon+P) \mathbf{u}}{v_F^2\tau} - \frac{\hbar \bm{\omega} n_5}{2 v_F \tau} ,
\end{eqnarray}
where $\eta$ and $\zeta$ are the shear and bulk viscosities (see, e.g., Ref.~\cite{Landau:t6}), $B\equiv|\mathbf{B}|$,
$\epsilon$ is the energy density, $P$ is the pressure, $\mathbf{u}$ is the electron fluid velocity,
$\bm{\omega}=\left[\bm{\nabla}\times\mathbf{u}\right]/2$ is the vorticity, and $v_F$ is the Fermi velocity.
Finally,  $-en$ and $-en_{5}$ are the matter parts of the electric and chiral charge densities, respectively.
Here, we use the convention that the derivatives act on all terms to their right. For the sake of simplicity,
we ignored the spatial dependence of the viscosity coefficients.
In general, the term $\bm{\nabla}P$ can be considered as a consequence of an external force
applied to a sample (see, e.g., Ref.~\cite{Levitov-Falkovich:2016}).
It is worth noting, however, that such a term in the self-consistent approach leads to the gradient of the chemical potentials $\bm{\nabla}\mu$ and $\bm{\nabla}\mu_5$,
as well as the temperature gradient $\bm{\nabla} T$.
In this case, the system of the hydrodynamic equations becomes interconnected, which significantly complicates the calculations.
Therefore, in our study, we neglect the effects connected with $\bm{\nabla}P$.

The anomalous transport coefficients are given by \cite{Gorbar:2017vph}:
\begin{eqnarray}
\label{model-sigma-CKT-be}
\sigma^{(B)} &=& \frac{e^2\mu_5}{2\pi^2\hbar^2c}, \qquad \sigma_5^{(B)} = \frac{e^2\mu}{2\pi^2\hbar^2c},\\
\sigma^{(\epsilon, u)} &=& \frac{e^2v_F}{120\pi^2\hbar c^2},  \qquad \sigma^{(\epsilon, B)} = -\frac{e\mu\mu_5}{2\pi^2\hbar^2v_F c}, \\
\sigma^{(V)} &=& -\frac{e\mu\mu_{5}}{\pi^2v_F^2\hbar^2}, \qquad \sigma_5^{(V)} = -\frac{e}{2\pi^2\hbar^2v_F^2}\left(\mu^2+\mu_5^2 + \frac{\pi^2T^2}{3}\right), \\
\sigma^{(\epsilon, V)} &=& -\frac{e\mu}{6\pi^2\hbar v_F}, \qquad \sigma^{(\epsilon, V)}_5 = -\frac{e\mu_5}{6\pi^2\hbar v_F},
\label{model-sigma-CKT-ee}
\end{eqnarray}
which agree with those obtained in Refs.~\cite{Son:2012wh,Landsteiner:2012kd,Stephanov:2015roa} in the ``no-drag" frame
\cite{Rajagopal:2015roa,Stephanov:2015roa,Sadofyev:2015tmb}. Here $\mu$ is the electric chemical potential, $\mu_5$ is the chiral chemical potential, and $T$ is temperature.

It might be important to comment on the last two terms on the right-hand side of Eq.~(\ref{model-Euler}).
They describe the scattering of electrons on impurities and/or phonons in the relaxation-time approximation.
The relaxation time $\tau$ is due to the intravalley (chirality preserving) scattering processes. In our study,
we ignore the chirality-flipping intervalley processes whose relaxation time $\tau_5$ is usually much larger
than $\tau$ (see, e.g., Ref.~\cite{Zhang-Xiu:2015}). Although the relaxation time, in general, depends on $\mu$, we assume that such a dependence is weak and treat $\tau$ as a constant (however, by using the experimental results in Ref.~\cite{Gooth:2017}, the dependence on $T$ is taken into account).
The corresponding simplification should not change our qualitative results for the hydrodynamic flow, albeit it could affect quantitative features. Let us point out that the penultimate term in Eq.~(\ref{model-Euler}) does not
contain any derivative of $\mathbf{u}$ and, consequently, breaks the Galilean invariance. This reflects the
existence of the preferred coordinate system connected with the stationary lattice of ions in a solid.

The energy conservation relation in the CHD \cite{Gorbar:2017vph} amended by the viscosity terms reads
\begin{eqnarray}
\label{model-energy}
&&-\eta \left(\mathbf{u}\Delta \mathbf{u}\right) -\left(\zeta+\frac{\eta}{3}\right) \left(\mathbf{u}\cdot\bm{\nabla}\right)\left(\bm{\nabla}\cdot\mathbf{u}\right) - \kappa\bm{\nabla}\left(\bm{\nabla} T - \frac{T}{\epsilon+P} \bm{\nabla}P\right)
+(\bm{\nabla}\cdot\mathbf{u}) (\epsilon+P) \nonumber\\
&& +\sigma^{(\epsilon, u)} \left[\sum_{i=1}^3B_i \left(\mathbf{B}\cdot\bm{\nabla}\right)u_i -B^2(\bm{\nabla}\cdot\mathbf{u})\right]-\frac{2c\left(\mathbf{E}\cdot\left[\bm{\nabla}\times\mathbf{u}\right]\right)\sigma^{(\epsilon,B)}}{3v_F}
-5c\sigma^{(\epsilon, u)}\left(\mathbf{E}\cdot\left[\bm{\nabla}\times\mathbf{B}\right]\right)
\nonumber\\
&&+v_F\left(\mathbf{B}\cdot\bm{\nabla}\right)\sigma^{(\epsilon, B)} +\frac{\hbar v_F (\bm{\nabla}\cdot\bm{\omega})n_5}{2}- \frac{\left(\mathbf{E}\cdot\left[\bm{\nabla} \times \bm{\omega}\right]\right)\sigma^{(\epsilon, V)}}{2} -2\sigma^{(\epsilon, u)}
\left[\left(\mathbf{u}\cdot\bm{\nabla}\right)B^2 -3\sum_{i=1}^3u_i\left(\mathbf{B}\cdot\bm{\nabla}\right) B_i\right] \nonumber\\
&&=-\left(\mathbf{E}\cdot \left[en\mathbf{u}-\sigma^{(B)}\mathbf{B} -\frac{\sigma^{(V)}\bm{\omega}}{3}
\right]\right).
\end{eqnarray}
The first two terms on the left-hand side describe the energy dissipation due to viscosity
\cite{Landau:t6}. The third term on the left-hand side of Eq.~(\ref{model-energy}) is related to the
thermoconductivity and is important for the self-consistency of the complete set of the CHD equations.
The corresponding coefficient is given by Eq.~(\ref{model-kappa}) in the main text and is assumed to be uniform.
By using the same assumptions as in Eq.~(\ref{model-Euler}), we also omit $\bm{\nabla}P$.

It is worth noting that the hydrodynamics equations (\ref{model-Euler}) and (\ref{model-energy}) were obtained
in Ref.~\cite{Gorbar:2017vph} from the consistent chiral kinetic equations \cite{Gorbar:2016ygi}
by averaging them with the quasiparticle momentum and energy \cite{Landau:t10,Huang-book}.

\section{Generated magnetic field, electric charge density, and the case of vanishing thermoelectric conductivity}
\label{sec:app-hydro-vars}

In this appendix, we present the general expressions for the generated magnetic field $\mathbf{B}(y)$, the electric charge density $n(y)$, as well as the solutions for the hydrodynamic flow in the special case of the vanishing thermoelectric conductivity,
$\kappa_e=0$.

\subsection{Generated magnetic field}
\label{sec:app-hydro-vars-MF}

It is straightforward to obtain the analytical expressions for the
components of the magnetic field from Eqs.~(\ref{LCHD-general-Bz}) and (\ref{LCHD-general-Bx}) in the main text.
This is achieved by using the electric field $E_y(y)$ as well as fluid flow velocities $u_x(y)$ and $u_y(y)$ for the parity-inversion (PI) symmetric case; see Eqs.~(\ref{LCHD-T-Bz0-Ey}), (\ref{LCHD-T-Bz0-ux}), and (\ref{LCHD-T-Bz0-uy-Sol}), respectively. The
corresponding results read
\begin{eqnarray}
\label{LCHD-T-Bz0-Bz-Sol}
B_z(y) &=&  \frac{4\pi \mu_m}{c} \frac{v_F^2 \tau e^2n^2 E_x}{\epsilon+P}\left[y- \gamma \frac{\sinh{\left(\lambda_x y -\lambda_xL_y/2\right)}}{\lambda_x\cosh{\left(\lambda_xL_y/2\right)}}\right] +\frac{4\pi \mu_m}{c} \sigma E_x y \nonumber\\
&-& \frac{2\mu_me^3 b_z}{\pi \hbar^2 c^2 \sigma} \left(\frac{e^3b_zE_x}{2\pi^2\hbar^2c} + \kappa_eC_T \right) \Bigg\{ \frac{en}{N \sigma} \left[en - \frac{\kappa_e(\epsilon+P)}{\kappa}\right] \left[y- \frac{\sinh{\left(\lambda_{y} y -
\lambda_{y}L_y/2\right)}}{\lambda_{y}\cosh{\left(\lambda_{y}L_y/2\right)}}\right]-y
\Bigg\} \nonumber\\
&-&\gamma \frac{\mu_m \pi \sigma^{(\epsilon, V)} v_F^2 \tau enE_x \lambda_{x}}{c(\epsilon+P)}
\frac{\sinh{\left(\lambda_{x} y -\lambda_{x}L_y/2\right)}}{\cosh{\left(\lambda_{x}L_y/2\right)}} +C_{B_z},\\
\label{LCHD-T-Bz0-Bx-Sol}
B_x(y) &=& -\frac{2e^3\mu_m b_x}{\pi \hbar^2 c^2\sigma}\left(\frac{e^3b_zE_x}{2\pi^2\hbar^2c} + \kappa_eC_T \right) \Bigg\{
\frac{en}{N \sigma} \left[en - \frac{\kappa_e(\epsilon+P)}{\kappa}\right] \left[y- \frac{\sinh{\left(\lambda_{y} y -\lambda_{y}
L_y/2\right)}}{\lambda_{y}\cosh{\left(\lambda_{y}L_y/2\right)}}\right]
- y\Bigg\}\nonumber\\
&+&\frac{2e^3\mu_m b_yE_x}{\pi \hbar^2 c^2} y
+C_{B_x},
\end{eqnarray}
where $\sigma$, $\kappa_e$, $\lambda_x$, $N$, and $\lambda_y$ are given in Eqs.~(\ref{model-sigma}), (\ref{model-kappae}), (\ref{LCHD-T-Bz0-Euler-x-Sol-lambda}), (\ref{LCHD-T-Bz0-tu-def}), and
(\ref{LCHD-T-Bz0-tlambday-def}), respectively. Further,
\begin{eqnarray}
\label{LCHD-T-Bz0-CBz-def}
C_{B_z} &=& B_z\left(\frac{L_y}{2}\right) -\frac{2\pi \mu_m L_y}{c} \frac{v_F^2 \tau e^2n^2 E_x}{\epsilon+P} -\frac{2\pi \mu_m L_y}{c} \sigma
E_x \nonumber\\
&+& \frac{\mu_m L_y e^3 b_z}{\pi \hbar^2 c^2 \sigma} \left(\frac{e^3b_zE_x}{2\pi^2\hbar^2c} + \kappa_eC_T \right)
\left\{ \frac{en}{N \sigma} \left[en - \frac{\kappa_e(\epsilon+P)}{\kappa}\right] -1
\right\},\\
\label{LCHD-T-Bz0-CBx-def}
C_{B_x} &=& B_x\left(\frac{L_y}{2}\right) +\frac{e^3\mu_m b_x L_y}{\pi \hbar^2 c^2 \sigma}\left(\frac{e^3b_zE_x}{2\pi^2\hbar^2c} +
\kappa_eC_T \right) \left\{ \frac{en}{N \sigma} \left[en - \frac{\kappa_e(\epsilon+P)}{\kappa}\right]
- 1\right\}
-\frac{e^3\mu_m b_yE_x L_y}{\pi \hbar^2 c^2}.
\end{eqnarray}
In view of the symmetry of the problem, the induced magnetic field vanishes in the middle of the slab.
This implies that the first terms in Eqs.~(\ref{LCHD-T-Bz0-CBz-def}) and (\ref{LCHD-T-Bz0-CBx-def}) equal zero.
(Strictly speaking, the symmetry arguments alone may not be sufficient to ensure that the magnetic field
vanishes in the middle of the slab when the time-reversal symmetry is broken. The self-consistent solution for the hydrodynamic flow confirms, however, that the symmetry arguments indeed hold and,
consequently, the field is zero at $y=L_y/2$.)

As one can check, the presence of $b_y$ and/or both $b_x$ and $b_z$ is crucial for the generation of the
magnetic field pointing in the $x$ direction [in other cases $B_x(y)=0$]. Indeed, it is determined by the
Chern--Simons currents that include nonhydrodynamic $\propto b_y E_x$ and hydrodynamic $\propto b_xE_y(y)$ terms.
According to Eq.~(\ref{LCHD-T-Bz0-Bx-Sol}), the latter requires both $b_x$ and $b_z$ to be nonzero.
Unlike $B_x(y)$, the component of the magnetic field $B_z(y)$ is always generated and is determined primarily
by the Ohmic and hydrodynamic currents.
Therefore, it is not surprising that $B_z(y)$ is different for the free-surface and no-slip boundary conditions.
On the other hand, $B_x(y)$ is determined only by the normal flow and the Chern--Simons current and, consequently, is completely
insensitive to the choice of the boundary conditions.

\subsection{Spatially inhomogeneous electron charge density}
\label{sec:app-dn}

The spatially inhomogeneous part of the electron charge density is obtained from Gauss's law (\ref{LCHD-mu5-Gauss}) with the electric field (\ref{LCHD-T-Bz0-Ey}) and equals
\begin{eqnarray}
\label{LCHD-T-Bz0-dn}
n(y) &=& n -\frac{\varepsilon_e}{4\pi e} \partial_y E_y(y) - \frac{e^2}{2\pi^2\hbar^2 c^2} \left[b_z B_z(y)+b_x B_x(y)\right]=
n -\frac{\varepsilon_e}{4\pi e \sigma} \left[en - \frac{\kappa_e(\epsilon+P)}{\kappa}\right] \partial_yu_y(y) \nonumber\\
&-& \frac{e^2}{2\pi^2\hbar^2 c^2} \left[b_z B_z(y)+b_x B_x(y)\right].
\end{eqnarray}
We checked that the relative deviation of the electric charge density $n(y)$ from $n$ is noticeable only
when $b_z\neq0$ and stems primarily from the Chern--Simons term, i.e., the last term in Eq.~(\ref{LCHD-T-Bz0-dn}).
As expected, the effect of the boundary conditions is relatively weak for large $L_y$ and strong for the small $L_y$.
In addition, $n(y)-n$ quickly diminishes with $\mu$ and $T$, but grows with the slab thickness $L_y$.

\subsection{The case of zero thermoelectric conductivity and preserved PI symmetry}
\label{sec:app-hydro-vars-kappae=0}

If the thermoelectric conductivity vanishes $\kappa_e=0$, then the transverse component of the electron flow velocity (\ref{LCHD-T-Bz0-uy-Sol}) can be simplified as
\begin{equation}
\label{LCHD-Bz0-uy-Sol}
u_y(y) = \frac{e^4nb_zE_x}{2\pi^2\hbar^2c\sigma N_0} \left[1- \frac{\cosh{\left(\lambda_{0,y} y -\lambda_{0,y}L_y/2\right)}}{\cosh{\left(\lambda_{0,y}L_y/2\right)}}\right],
\end{equation}
where
\begin{eqnarray}
\label{LCHD-Bz0-U-def}
N_0 &=& \frac{e^2n^2}{\sigma} + \frac{\epsilon+P}{v_F^2\tau},\\
\label{LCHD-Bz0-lambday-def}
\lambda_{0,y} &=& \sqrt{\frac{e^2n^2}{\eta_{y} \sigma} + \frac{\epsilon+P}{\eta_{y} v_F^2\tau}},
\end{eqnarray}
and $\eta_{y}\equiv4\eta/3$.
The corresponding electric field $E_y(y)$ reads
\begin{equation}
\label{LCHD-Bz0-Ey}
E_y(y) = \frac{en}{\sigma} u_y(y)  - \frac{e^3 E_x b_z}{2\pi^2 \hbar^2 c \sigma}
= \frac{e^5n^2b_zE_x}{2\pi^2\hbar^2c\sigma N_0} \left[1- \frac{\cosh{\left(\lambda_{0,y} y -\lambda_{0,y}L_y/2\right)}}{\cosh{\left(\lambda_{0,y}L_y/2\right)}}\right] - \frac{e^3 E_x b_z}{2\pi^2 \hbar^2 c \sigma}.
\end{equation}

Equations (\ref{LCHD-T-Bz0-Bz-Sol}) and (\ref{LCHD-T-Bz0-Bx-Sol}) are also simplified and read as
\begin{eqnarray}
\label{LCHD-Bz0-Bz-Sol}
B_z(y) &=& \frac{4\pi \mu_m}{c}\Bigg\{ \frac{v_F^2 \tau e^2n^2 E_x}{\epsilon+P}\left[y- \gamma \frac{\sinh{\left(\lambda_x y -\lambda_xL_y/2\right)}}{\lambda_x\cosh{\left(\lambda_xL_y/2\right)}}\right] +\sigma E_x y \nonumber\\
&+& \frac{E_x}{\sigma} \left(\frac{e^3b_z}{2\pi^2 \hbar^2 c}\right)^2 \left[y -\frac{e^2n^2}{\sigma N_0}\left(y- \frac{\sinh{\left(\lambda_{0,y} y -\lambda_{0,y}L_y/2\right)}}{\lambda_{0,y}\cosh{\left(\lambda_{0,y}L_y/2\right)}}\right) \right] \Bigg\} \nonumber\\
&-& \gamma \frac{\mu_m \pi \sigma^{(\epsilon, V)} v_F^2 \tau enE_x \lambda_{x}}{c(\epsilon+P)}
\frac{\sinh{\left(\lambda_{x} y -\lambda_{x}L_y/2\right)}}{\cosh{\left(\lambda_{x}L_y/2\right)}} +C_{B_z},\\
\label{LCHD-Bz0-Bx-Sol}
B_x(y) &=& -\frac{2e^3\mu_m}{\pi \hbar^2 c^2}\Bigg\{ \frac{e^5n^2b_xb_zE_x}{2\pi^2\hbar^2c\sigma^2N_0} \left[y- \frac{\sinh{\left(\lambda_{0,y} y -\lambda_{0,y}L_y/2\right)}}{\lambda_{0,y}\cosh{\left(\lambda_{0,y}L_y/2\right)}}\right]
-\frac{e^3E_x b_xb_z}{2\pi^2\hbar^2c\sigma} y - b_yE_xy \Bigg\}
+C_{B_x},
\end{eqnarray}
where
\begin{eqnarray}
\label{LCHD-Bz0-CBz-def}
C_{B_z} &=& B_z\left(\frac{L_y}{2}\right) - \frac{ 2\pi \mu_mL_y}{c}\left[\frac{e^2n^2v_F^2 \tau E_x}{\epsilon+P} +\sigma E_x +\frac{E_x}{\sigma} \left(\frac{e^3b_z}{2\pi^2\hbar^2c}\right)^2\left(1-\frac{e^2n^2}{\sigma N_0}\right)\right],\\
\label{LCHD-Bz0-CBx-def}
C_{B_x} &=& B_x\left(\frac{L_y}{2}\right) + \frac{e^3\mu_mL_y}{\pi\hbar^2c^2}\left[\frac{e^5n^2 b_xb_zE_x}{2\pi^2\hbar^2c\sigma^2N_0} - \frac{e^3E_xb_xb_z}{2\pi^2\hbar^2c\sigma} -b_yE_x\right].
\end{eqnarray}

\end{document}